\documentclass[aps,prd,groupaddress,tighten,nofootinbib,preprint,superscriptaddress]{revtex4}
\usepackage{pifont}
\usepackage{amssymb}
\usepackage{amsmath,color}
\usepackage{epsfig}
\usepackage{graphicx,epsf,epsfig}
\usepackage{mathrsfs}
\usepackage{bbm}
\def\slc#1{\setbox0=\hbox{$#1$}           
    \dimen0=\wd0                                 
    \setbox1=\hbox{/} \dimen1=\wd1               
    \ifdim\dimen0>\dimen1                        
       \rlap{\hbox to \dimen0{\hfil/\hfil}}      
       #1                                        
    \else                                        
       \rlap{\hbox to \dimen1{\hfil$#1$\hfil}}   
       /                                         
    \fi}

\newcommand{\bi}{\begin{itemize}}
\newcommand{\ei}{\end{itemize}}

\newcommand{\be}{\begin{equation}}
\newcommand{\ee}{\end{equation}}
\newcommand{\bea}{\begin{eqnarray}}
\newcommand{\eea}{\end{eqnarray}}

\newcommand{\ldm}{\Delta m_{31}^2}
\newcommand{\sdm}{\Delta m_{21}^2}

\newcommand{\stheta}{\sin^2 2 \theta_{13}}

\newcommand{\msc}[1]{\mathscr{#1}}
\newcommand{\trm}[1]{\textrm{#1}}
\newcommand{\mcl}[1]{\mathcal{#1}}

\newcommand{\ie}{{\it i.e.}}

\newcommand{\eg}{{\it e.g.}}

\newcommand{\cf}{{\it cf.}}

\newcommand{\eq}{Eq.}

\newcommand{\fig}{Fig.}

\newcommand{\Ref}{Ref.}
\newcommand{\Refs}{Refs.}
\newcommand{\Sec}{Sec.}

\newcommand{\Tab}{Table}

\newcommand{\equ}[1]{\eq~(\ref{equ:#1})}
\newcommand{\figu}[1]{\fig~\ref{fig:#1}}

\begin{document}
\preprint{NORDITA-2009-79} \preprint{EURONU-WP6-09-14}
\preprint{IDS-NF-014}

\title{Non-standard interactions versus non-unitary lepton flavor mixing at a neutrino factory}

\author{Davide Meloni}
\email{davide.meloni@physik.uni-wuerzburg.de}

\affiliation{Institut f{\"u}r Theoretische Physik und Astrophysik,
Universit{\"a}t W{\"u}rzburg, 97074 W{\"u}rzburg, Germany}

\author{Tommy Ohlsson}
\email{tommy@theophys.kth.se}

\affiliation{Department of Theoretical Physics, School of
Engineering Sciences, Royal Institute of Technology (KTH) --
AlbaNova University Center, Roslagstullsbacken 21, 106 91 Stockholm,
Sweden}

\author{Walter Winter}
\email{winter@physik.uni-wuerzburg.de}

\affiliation{Institut f{\"u}r Theoretische Physik und Astrophysik,
Universit{\"a}t W{\"u}rzburg, 97074 W{\"u}rzburg, Germany}

\author{He Zhang}
\email{zhanghe@kth.se}

\affiliation{Department of Theoretical Physics, School of
Engineering Sciences, Royal Institute of Technology (KTH) --
AlbaNova University Center, Roslagstullsbacken 21, 106 91 Stockholm,
Sweden}

\begin{abstract}
The impact of heavy mediators on neutrino oscillations is typically
described by non-standard four-fermion interactions (NSIs) or
non-unitarity (NU). We focus on leptonic dimension-six effective
operators which do not produce charged lepton flavor violation.
These operators lead to particular correlations among neutrino
production, propagation, and detection non-standard effects. We
point out that these NSIs and NU phenomenologically lead, in fact,
to very similar effects for a neutrino factory, for completely
different fundamental reasons. We discuss how the parameters and
probabilities are related in this case, and compare the
sensitivities. We demonstrate that the NSIs and NU can, in
principle, be distinguished for large enough effects at the example
of non-standard effects in the $\mu$-$\tau$-sector, which basically
corresponds to differentiating between scalars and fermions as heavy
mediators as leading order effect. However, we find that a near
detector at superbeams could provide very synergistic information,
since the correlation between source and matter NSIs is broken for
hadronic neutrino production, while NU is a fundamental effect
present at any experiment.
\end{abstract}
\maketitle

\section{Introduction}

During the past decade, experimental studies of neutrino
oscillations have provided us with compelling evidence that
neutrinos are massive particles and lepton flavors mix. Since an
important new window is opened for searching new physics beyond the
Standard Model (SM) of particle physics, it is interesting to
discuss the impact of potential non-standard effects\footnote{Note
that we will distinguish between the two concepts of non-standard
four-fermion interactions (NSIs) and non-unitarity (NU). These two
concepts will collectively be called non-standard effects.} on
neutrino oscillations. In this study, we focus our attention on
non-standard effects from heavy mediators, which are integrated out
at the scales of the neutrino oscillation experiments.

It is convenient to parameterize the impact of the heavy fields,
present in high-energy theory,  by adding a tower  of effective
operators $\mathcal{O}^{d}$ of dimension $d>4$ to the Lagrangian.
These {\it non-renormalizable} operators are made out of the SM
fields, and invariant under the SM gauge
group~\cite{Weinberg:1979sa,Wilczek:1979hc,Buchmuller:1985jz}. They
parameterize the effects of the high-energy degrees of freedom on the
low-energy theory order by order. In principle, the operator
coefficients are weighted by inverse powers of the scale of new
physics $\Lambda_{\mathrm{NP}}$:
 \begin{equation}\label{equ:leff}
\mathscr{L} = \mathscr{L}_{\rm SM} + \mathscr{L}^{d=5}_{\text{eff}}
+ \mathscr{L}^{d=6}_{\text{eff}} + \cdots \, , \quad \textrm{with}
\quad \mathscr{L}^{d}_{\text{eff}} \propto
\frac{1}{\Lambda_{\mathrm{NP}}^{d-4}} \, \mathcal{O}^{d}\,.
\end{equation}
The only possible dimension-five operator, namely,
$\msc{L}^{d=5}_{\trm{eff}}$, which violates lepton number by two
units, is the famous Weinberg operator~\cite{Weinberg:1979sa} \be
\mcl{O}^5_W = (\overline{L^{c}} {\rm i} \tau^{2} \phi)\, (\phi {\rm
i} \tau^2 L) \, , \label{equ:weinberg} \ee which leads, after
electroweak symmetry breaking (EWSB), to Majorana masses for the
neutrinos. Here $L$ and $\phi$ stand for the Standard Model lepton
doublets and the Higgs field, respectively. At tree level,
$\mcl{O}^5_W$ can only be mediated by a singlet fermion, a triplet
scalar, or a triplet fermion, leading to the famous
type~I~\cite{Minkowski:1977sc,Yanagida:1979as,GellMann:1980vs,Mohapatra:1979ia},
type~II~\cite{Magg:1980ut,Schechter:1980gr,Wetterich:1981bx,Lazarides:1980nt,Mohapatra:1980yp,Cheng:1980qt},
or type~III~\cite{Foot:1988aq} seesaw mechanism, respectively (see
also \Refs~\cite{Abada:2007ux,delAguila:2008pw}). Compared to the
electroweak scale, the masses of the neutrinos in all three cases
appear suppressed by a factor $v/\Lambda_{\mathrm{NP}}$, where
$v/\sqrt{2}$ is the vacuum expectation value (VEV) of the Higgs
field. Substituting typical values, one obtains that the original
seesaw mechanisms point towards the GUT scale.

Except for neutrino masses, the dimension-six operators potentially
affecting neutrino oscillations are, for non-standard four-fermion
interactions (NSIs),  operators of the types
\begin{equation}
{\cal O^S} = (\bar{E} E ) (\bar L L) \, , \quad (\bar{L} L ) (\bar L
L) \, , \label{equ:opnsi}
\end{equation}
and, for non-unitarity (NU) coming from heavy singlet fermions, of
the type
\begin{eqnarray}
{\cal O^F} = \left(\overline{L}
 \phi\right) {\rm i} \slc\partial \left(\phi^\dagger
L \right) \, , \label{equ:LNU}
\end{eqnarray}
where we omitted flavor, spin, and gauge indices.
Some of these effective operators result in corrections to the
low-energy SM parameters and in exotic couplings. For instance,
\equ{LNU} implies a correction to the neutrino kinetic energy. After
re-diagonalizing and re-normalizing the neutrino kinetic
terms~\cite{DeGouvea:2001mz,Broncano:2003fq}, a non-unitary leptonic
mixing matrix appears~\cite{Antusch:2006vwa,Abada:2007ux}. On the
other hand, \equ{opnsi} typically leads to lepton-flavor-violating
processes.

In ordinary seesaw models, the operators generating neutrino masses
[such as \equ{weinberg}] and non-standard effects [\cf,
Eqs.~\eqref{equ:opnsi} and \eqref{equ:LNU}] are both mediated by the
same heavy seesaw particles, and therefore, there might be
connections between them in some cases, in particular, in the
type-II seesaw. However, non-standard effects are usually suppressed
dramatically by the scale of the seesaw threshold, which is
typically not far away from the GUT scale. In some low-scale seesaw
models, the smallness of neutrino masses is protected by other
suppression mechanisms rather than the GUT scale, such as radiative
generation~\cite{Zee:1980ai,Wolfenstein:1980sy,Zee:1985rj,Babu:1988ki,Ma:1998dn,Babu:2002uu,Krauss:2002px,Cheung:2004xm,Ma:2006km,Ma:2007yx,Aoki:2008av,Aoki:2009vf},
small lepton number breaking
\cite{Schechter:1981bd,Nandi:1985uh,Mohapatra:1986bd,Branco:1988ex,GonzalezGarcia:1988rw,Ma:2000cc,Tully:2000kk,Loinaz:2003gc,Hirsch:2004he,Pilaftsis:2005rv,Gouvea:2007xp,Kersten:2007vk,Malinsky:2008qn,Grimus:2009mm,Malinsky:2009gw,Malinsky:2009df,Dev:2009aw,Zhang:2009ac},
or neutrino masses from higher-than-dimension-five effective
operators~\cite{Babu:1999me,Chen:2006hn,Gogoladze:2008wz,Giudice:2008uua,Babu:2009aq,Gu:2009hu,Bonnet:2009ej}.
In these cases, singlet mediators may be introduced at the TeV
scale, which typically lead to observable NU effects as well. As
another example, in \Tab~2 of \Ref~\cite{Bonnet:2009ej}, a number of
possibilities to generate small neutrino masses together with NU are
listed, where the neutrino mass originates from a dimension-seven
operator. Of course, the heavy seesaw particles may also be directly
searched for at future colliders, in particular, at the Large Hadron
Collider via the lepton-flavor-violating decays
\cite{delAguila:2007em}.

The NSI operators in \equ{opnsi} are typically connected to the
charged lepton flavor violation by SU(2) gauge invariance, and
constrained by lepton universality tests. This implies that
stringent bounds exist for the NSIs (we list the current bounds in
\Sec~\ref{sec:bounds}).  On the other hand, if one has a theory for
which the (gauge invariant) NSIs in \equ{opnsi} appear without
charged lepton flavor violation, the NSIs present in neutrino
production at a neutrino factory and the neutrino propagation are
correlated in a particular way, and the flavor structure of the NSIs
is strongly constrained~\cite{Gavela:2008ra}. Therefore, we focus on
this class of operators in the following. The most prominent example
for such a theory is the exchange of a heavy charged SU(2) singlet
scalar, leading to an effective operator of the type
\begin{equation}
\mathcal{O^S} = \left(\overline{L^c} \cdot L
\right)\left(\overline{L} \cdot L^c \right) \,
,\label{equ:exampleop}
\end{equation}
where the dot denotes the SU(2) invariant product, i.e., ${\rm i}
\tau^2$. This operator is antisymmetric in the flavor indices, and
does not lead to charged lepton flavor violation. In fact,
Eqs.~(\ref{equ:LNU}) and~(\ref{equ:exampleop}) are the only two
dimension-six operators which lead to non-standard effects without
charged lepton flavor violation.\footnote{In principle, there could
be four-fermion interactions with two quarks and two leptons
contributing to neutrino detection or matter effects at a neutrino
factory. However, it can be shown that these cannot produce NSIs
without charged lepton flavor violation if written in a
gauge-invariant way, because the up and down quark contributions
carry different coefficients which makes it impossible to cancel
both contributions simultaneously (without canceling the NSIs). See,
for example, Ref.~\cite{OtaTalk} for details.} In the remaining
parts of this work, we will use NSIs and NU to denote the
non-standard effects coming from Eqs.~\eqref{equ:exampleop} and
\eqref{equ:LNU}, respectively.

In fact, the above mentioned correlation between source and matter
NSIs at the neutrino factory increases the experimental sensitivity
to the NSI parameters dramatically~\cite{Tang:2009na}, similar to
the NU case~\cite{Antusch:2009pm}. Furthermore, it is well known
that NU can be re-parameterized in terms of
NSIs~\cite{FernandezMartinez:2007ms}. We will demonstrate that for a
neutrino factory, the mentioned NSIs are phenomenologically very
similar to NU, but for completely different fundamental reasons,
which makes it hard to distinguish them.  On the other hand, at tree
level, the natural implementation for the NSIs are the scalar bosons
leading to the operator in \equ{exampleop}, whereas the NU is
mediated by SM singlet fermions. Therefore, distinguishing between
the NSI and NU operators without charged lepton flavor violation is
basically equivalent to differentiating between scalars and fermions
as mediators, at least to leading order at tree level, and therefore
theoretically very interesting. Note that, besides leptonic NSIs,
there may exist non-standard neutrino-quark interactions stemming
from some Grand Unified or $R$-parity violating supersymmetric
theories which could also affect neutrino oscillations, but such
scenarios lie beyond the scope of this work.

The NSIs and NU have been extensively studied in the literature,
from both theoretical and phenomenological point of view. In
particular, it has been pointed out that the sub-leading effects
generated by
NSIs~\cite{Valle:1987gv,Guzzo:1991hi,Roulet:1991sm,Grossman:1995wx}
add to the standard matter effect
\cite{Wolfenstein:1977ue,Mikheev:1986gs} and also introduce new
sources of CP violation; then a neutrino factory or a superbeam
experiments are adequate places to study their effects
~\cite{NSInfstart,Ota:2001pw,GonzalezGarcia:2001mp,Gago:2001xg,Huber:2001zw,Kopp:2007mi,
Ribeiro:2007ud,Kopp:2008ds,Winter:2008eg,NSInfstop,NSIotherstart,Adhikari:2006uj,Blennow:2007pu,Kopp:2007ne,Blennow:2008ym,NSIotherstop}.
Since the sub-leading effects presented in the NU framework are
quite similar, the same future facilities can be used to constrain
(or measure) the additional rotations and phases
\cite{Antusch:2006vwa,FernandezMartinez:2007ms,Winter:2008eg,Antusch:2009pm,Malinsky:2009df,Altarelli:2008yr,Malinsky:2009gw}.

In this work, we are mainly interested in studying the experimental
signatures from both types of non-standard operators and  trying to
understand whether it is possible to disentangle them by using a
neutrino factory facility or not. To this end, we will first
present, in Sec.~\ref{sec:formulism}, the general formalism
depicting neutrino oscillations in matter with both source and
detector effects being included. These formulas are
model-independent, and could be used in theories with both a unitary
leptonic mixing matrix or a non-unitary one. In
Sec.~\ref{sec:operators}, we will then classify the
higher-dimensional operators according to the mediators, and figure
out the NSI effects induced by different non-renormalizable
operators. We also summarize the current bounds on the non-standard
parameters discussed in this work. In Sec.~\ref{sec:test}, we
briefly discuss several possibilities how to determine the origin of
the non-standard effects. Section~\ref{sec:last} is devoted to a
detailed analytical discussion of the transition probabilities
useful for our analysis, as well as to the presentation of our
simulation techniques and the numerical results to show the
prospects of searching for the origin of NSIs in a future neutrino
factory. Finally, a brief summary is given in
Sec.~\ref{sec:summary}.

\section{Neutrino oscillations with non-standard effects}
\label{sec:formulism}

In this section, we describe neutrino oscillations with non-standard
effects from heavy mediators. Note that, NU essentially affects the
couplings of neutrinos to gauge bosons $W$ and $Z$, which are
actually integrated out when describing neutrino oscillations, in
particular, in the presence of matter effects. Effectively, NU leads
to four-fermion interactions similar to these from NSIs, and
therefore, both NSIs and NU can be described by using the same
parametrization. For this part, it will be useful to treat both
classes within the NSI framework, since the source, propagation, and
detection effects are a {\it priori} treated independently for NSIs.
Therefore, we use the NSI parametrization in this section. Namely,
we present the oscillation probabilities including any type of
non-standard effects in term of NSI parameters
$\varepsilon_{\alpha\beta}$ in despite of their possible origins.

In order to perform more precision measurements on neutrino mixing
parameters, an intense high-energy neutrino source together with a
long-baseline setup is proved to be the best choice
\cite{Bandyopadhyay:2007kx}. Similar to the standard matter effects
in long-baseline neutrino oscillation experiments, NSIs can affect
the neutrino propagation by coherent forward scattering in Earth
matter. In the language of effective Hamiltonians, the time
evolution of neutrino flavor eigenstates in the presence of NSIs is
described by
\begin{eqnarray}
{\cal H} = {\cal H}_0 + {\cal H}_m + {\cal H}_{\rm NSI} \, ,
\label{equ:H}
\end{eqnarray}
where
\begin{eqnarray}
{\cal H}_0 & = &  \frac{1}{2E}U {\rm diag} (m^2_1,m^2_2,m^2_3)
U^\dagger  \, ,  \label{eq:H0}\\
{\cal H}_m & = & {\rm diag} (V_{\rm CC},0,0) \, , \\
{\cal H}_{\rm NSI} & = & V_{\rm CC} \varepsilon^m \, ,
\end{eqnarray}
with $V_{\rm CC} \simeq \sqrt{2}G_F N_e $ arising from coherent
forward scattering and $N_e$ denoting the electron number density
along the neutrino trajectory in Earth. The vacuum leptonic mixing
matrix $U$ is usually parameterized in the standard form by using
three mixing angles and one CP violating phase
\begin{eqnarray}\label{eq:para}
U & = &\left(
\begin{matrix}c_{12} c_{13} & s_{12} c_{13} & s_{13}
e^{-{\rm i}\delta} \cr -s_{12} c_{23}-c_{12} s_{23} s_{13}
 e^{{\rm i} \delta} & c_{12} c_{23}-s_{12} s_{23} s_{13}
 e^{{\rm i} \delta} & s_{23} c_{13} \cr
 s_{12} s_{23}-c_{12} c_{23} s_{13}
 e^{{\rm i} \delta} & -c_{12} s_{23}-s_{12} c_{23} s_{13}
 e^{{\rm i} \delta} & c_{23} c_{13}\end{matrix}
\right) \, ,
\end{eqnarray}
where $c_{ij} \equiv \cos \theta_{ij}$, $s_{ij} \equiv \sin
\theta_{ij}$ (for $ij=12$, $13$, $23$), and $\delta$ is the Dirac
CP-violating phase. In analogy to the vacuum Hamiltonian ${\cal
H}_0$ in Eq.~\eqref{eq:H0}, the effective Hamiltonian ${\cal H}$ can
also be diagonalized through a unitary transformation
\begin{eqnarray}\label{eq:Heff}
{\cal H} = \frac{1}{2E} \tilde{U} {\rm diag}\left( \tilde{m}^2_1,
\tilde{m}^2_2, \tilde{m}^2_3 \right) \tilde{U}^{\dagger} \, ,
\end{eqnarray}
where $\tilde{m}^{2}_i$ ($i=1,2,3$) denote the effective mass
squared eigenvalues of neutrinos and $\tilde U$ is the effective
leptonic mixing matrix in matter. Note that, in writing down
Eq.~\eqref{eq:Heff}, we have already taken into account the
Hermitian property of ${\cal H}$. The explicit expressions for
$\tilde{U}$ and $\tilde m^2_i$ can be found in
Ref.~\cite{Meloni:2009ia}.

In addition to propagation in matter, production or detection
processes can be affected by NSIs. The neutrino states produced in a
source and observed at a detector can be treated as superpositions
of pure orthonormal flavor states \cite{Grossman:1995wx,
Gonzalez-Garcia:2001mp, Bilenky:1992wv}:
\begin{eqnarray}\label{eq:s}
|\nu^s_\alpha \rangle & = &  |\nu_\alpha \rangle +
\sum_{\beta=e,\mu,\tau} \varepsilon^s_{\alpha\beta}
|\nu_\beta\rangle  = (1 + \varepsilon^s) U  |\nu_m \rangle \ , \\
\langle \nu^d_\beta| & = &\langle
 \nu_\beta | + \sum_{\alpha=e,\mu,\tau}
\varepsilon^d_{\alpha \beta} \langle  \nu_\alpha  | =  \langle \nu_m
|  U^\dagger (1 + (\varepsilon^d)^\dagger) \ , \label{eq:d}
\end{eqnarray}
where the superscripts `$s$' and `$d$' denote the source and the
detector, respectively. Note that the two flavor indices follow from
the postulate of the coherent contribution to the source or
detection effects, and more generally, there could be incoherent
contributions, which will not be considered further
\cite{Gonzalez-Garcia:2001mp,Huber:2002bi}.

For NSIs, the parameters at sources and detectors are not
necessarily correlated. Only if the production and the detection are
exactly the same process with the same other participating fermions
(e.g., beta decay and inverse beta decay), the same quantity enters
as $\varepsilon_{\alpha \beta}^s=(\varepsilon_{\beta
\alpha}^d)^*$~\cite{Kopp:2007ne}. It is important to keep in mind
that these NSI parameters are experiment- and process-dependent
quantities. In the following, we will mainly focus on the source and
detector NSIs defined for a neutrino factory and specific processes.
For NU, however, source, propagation, and detection effects are
correlated in a particular way, as we will discuss in the next
section.

Including all the NSI effects into the neutrino oscillations,
we arrive at the amplitude for the process $\nu^s_\alpha
\rightarrow \nu^d_\beta$
\begin{eqnarray}\label{eq:effA1}
{\cal A}_{\alpha\beta}(L) & = & \langle \nu^d_\beta | {\rm e}^{-{\rm
i} {\cal H} L} |\nu^s_\alpha \rangle = (1 +
{\varepsilon^d})_{\rho\beta} { A}_{\gamma\rho}\left(1 +
{\varepsilon^s}\right)_{\alpha\gamma} \nonumber \\
&=& \left[(1 + {\varepsilon^d})^T { A}^T\left(1 +
{\varepsilon^s}\right)^T\right]_{\beta\alpha} = \left[ { A} +
{\varepsilon^s} { A} + { A} {\varepsilon^d} + {\varepsilon^s} { A}
{\varepsilon^d} \right]_{\alpha \beta} \, ,
\end{eqnarray}
where $L$ is the propagation distance and $A$ is a coherent sum over
the contributions of all the mass eigenstates $\nu_i$
\begin{eqnarray}\label{eq:A}
{A}_{\alpha\beta} = \sum_i \tilde U^*_{\alpha i} \tilde U_{\beta i}
{\rm e}^{-{\rm i} \frac{\tilde m^2_i L}{2E}}\, .
\end{eqnarray}
With the above definitions, the oscillation probability is given by
\cite{Ohlsson:2008gx}
\begin{eqnarray}\label{eq:P1}
P(\nu^s_\alpha \rightarrow \nu^d_\beta) &=& \left| {\cal
A}_{\alpha\beta}(L) \right|^2\
\nonumber \\
& = & \sum_{i,j} {\cal J}^i_{\alpha\beta} {\cal
J}^{j*}_{\alpha\beta} - 4 \sum_{i>j} {\rm Re} ({\cal
J}^i_{\alpha\beta} {\cal J}^{j*}_{\alpha\beta} )\sin^2\frac{\Delta
\tilde m^2_{ij}L}{4E} \nonumber \\
&&+ 2 \sum_{i>j}{\rm Im} ( {\cal J}^i_{\alpha\beta} {\cal
J}^{j*}_{\alpha\beta} ) \sin\frac{ \Delta \tilde m^2_{ij} L}{2 E} \,
,
\end{eqnarray}
where
\begin{eqnarray}\label{eq:J}
{\cal J}^i_{\alpha\beta}  &=& {\tilde U^*_{\alpha i} \tilde U_{\beta
i} + \sum_\gamma \varepsilon^s_{\alpha \gamma} \tilde U^*_{\gamma i}
\tilde U_{\beta i}+ \sum_\gamma \varepsilon^d_{\gamma \beta} \tilde
U^*_{\alpha i} \tilde U_{\gamma i}     + \sum_{\gamma,\rho}
\varepsilon^s_{\alpha\gamma} \varepsilon^d_{\rho \beta} \tilde
U^*_{\gamma i} \tilde U_{\rho i} }\, .
\end{eqnarray}
A salient feature of Eq.~\eqref{eq:P1} is that, when
$\alpha\neq\beta$, the first term in Eq.~\eqref{eq:P1} is, in
general, not vanishing, and therefore, a flavor transition would
already happen at the source even before the oscillation process and
is known as the zero-distance effect \cite{Langacker:1988up}.
Although the effective leptonic mixing matrix in matter $\tilde U$
is still unitary, the presences of NSIs in the source and the
detector prevent us from defining a unique CP invariant quantity
like the standard Jarlskog invariant \cite{Jarlskog:1985ht}. New CP
non-conservation terms, which are proportional to the NSI parameters
and have different dependences on the quantity $L/E$, will appear in
the oscillation probability. Another peculiar feature in the
survival probability is that, in the case of $\alpha=\beta$,
CP-violating terms in the last line of Eq.~\eqref{eq:P1} may, in
principle, not vanish. Note that Eq.~\eqref{eq:P1} is also valid in
the case of a non-unitary leptonic mixing matrix. In the minimal
unitarity violation scheme, the NU effects are parameterized by
using similar $\varepsilon$ parameters as in the case of the source
and detector NSI effects, but with the relation $\varepsilon_{\alpha
\beta}^s=\varepsilon_{\alpha \beta}^d=(\varepsilon_{\beta
\alpha}^s)^*=(\varepsilon_{\beta \alpha}^d)^*$.

In what follows, we will discuss the possible origin of NSIs in
Eq.~\eqref{eq:P1} together with the correlations among the NSI
parameters.

\section{Origin of non-standard effects}
\label{sec:operators}

As mentioned in the Introduction, non-standard effects naturally
emerge from most fundamental theories beyond the SM, and can, in
general, be described by a series of higher-dimensional
non-renormalizable operators after integrating out the heavy degrees
of freedom in the underlying theory. The only dimension-five (${\cal
O}^5$) operator is the well-known Weinberg operator in
\equ{weinberg}, which gives birth to the masses of light neutrinos.
For dimension-six operators (${\cal O}^6$), depending on different
mediators, there are, in general, two different kinds of
non-renormalizable operators responsible for non-standard effects at
tree-level. For scalar mediated dimension-six operators ${\cal
O^S}$, four-fermion interactions are involved, which usually break
the lepton flavor (and lepton universality), but conserve the
unitarity of the leptonic mixing matrix. On the other hand, fermion
mediated dimension-six operators ${\cal O^F}$ correct the kinetic
energy terms of light neutrinos, which violate the unitarity of the
leptonic mixing matrix as a consequence of the mixing between light
and heavy neutral fermions. As for dimension-eight or higher
operators, both two types of non-standard effects can be induced. In
the following, we will discuss the possible non-standard effects
stemming from different kinds of higher-dimensional operators.

\subsection{Dimension-six operators mediated by scalars}
\label{sec:scalarvector}

If new scalars are introduced, the (leptonic) dimension-six NSI
operators mediated by these at tree level below the EWSB scale are
usually given by
\cite{Buchmuller:1985jz,Bergmann:1998ft,Bergmann:1999pk,Berezhiani:2001rs}
\begin{eqnarray}
{\cal O^S} = 2\, \sqrt{2} \, G_F \, (\varepsilon^{L/R})^{\alpha
\gamma}_{\beta \delta} \,
 \left(
 \bar{\nu}^{\beta} \gamma^{\rho} {\rm P}_{L} \nu_{\alpha}
  \right) \,
 \left(
 \bar{\ell}^{\delta} \gamma^{\rho} {\rm P}_{L/R} \ell_{\gamma}
 \right) \, ,
\label{equ:nsi}
\end{eqnarray}
where $\ell$ denote the charged leptons. Here $G_F$ is the Fermi
coupling constant and $P_L$ and $P_R$ are the left- and right-handed
(chiral) projection operators, respectively. Note that, in writing
down Eq.~\eqref{equ:nsi}, we do not require gauge invariance. If
SU(2) gauge invariance is imposed at the effective operator level
and it is required that all the charged-lepton processes vanish,
only the NSI operators of left-handed fields [the second type in
\equ{opnsi}] survive, which are antisymmetric in the flavor indices,
i.e., $ \alpha \neq \gamma$ and $\beta \neq \delta$. Such operators
can be naturally realized in theories with an SM SU(2) singlet
singly charged scalar, leading to the operator in
\equ{exampleop}~\cite{Bergmann:1999pk,Bilenky:1993bt,Cuypers:1996ia,Antusch:2008tz,Ohlsson:2009vk}.\footnote{However,
if appropriate cancellation conditions apply, one can also have
theories without charged lepton flavor violation with triplet
scalars, and singlet and triplet vectors. This can be read off from
\Tab~2 in \Ref~\cite{Gavela:2008ra}: the effective operators
generated by the exchange of the different mediators have to be
combined such that
$\mathcal{C}_{LL}^{\mathbf{1}}=-\mathcal{C}_{LL}^{\mathbf{3}}$. In
addition, there may be loop-induced dimension-six operators.  Since
in these cases the theoretical interpretation is obscure (there
maybe other effects induced by these operators), we focus on the
singlet scalar interpretation in the following, which is the
simplest one.} Therefore, the observation of a dimension-six
operator of this type in the neutrino sector may be interpreted in
terms of a heavy scalar boson as mediator. Note that, conversely,
the consequence of any theory which does not lead to
charged-lepton-flavor-violation and produces dimension-six operators
is, in general, that the antisymmetric conditions must
hold~\cite{Gavela:2008ra}.

For a neutrino factory, the (leptonic) NSI effects induced by
Eq.~\eqref{equ:nsi} are relevant for the source, but not for the
detector, since the detection processes involve quarks. Compared
with Eqs.~\eqref{eq:s} and \eqref{eq:d}, one can easily read
off\footnote{Note that there is no standard $\nu_\tau$ production in
a neutrino factory, and hence, there is no corresponding NSI
parameter like $\varepsilon^s_{\tau\beta}$.}
\begin{eqnarray}
\varepsilon^{\mathrm{NF}}_{e \beta} & = & (\varepsilon^L)^{\mu
e}_{\beta \mu} \, ,
\\
\varepsilon^{\mathrm{NF}}_{\mu \beta} & = & (\varepsilon^L)^{e
\mu}_{\beta e} \, .
\end{eqnarray}
Here the dominating effect comes from the left-handed component at
the production process due to the helicity suppression of the
right-handed component.  Note that we use the label ``NF'' to mark
that these NSIs are (production) process dependent quantities only
relevant for a neutrino factory, whereas potential NSIs at a
superbeam are, in general, completely uncorrelated. In addition,
note that by giving up two of the four flavor indices, these
parameters violate CP and even CPT explicitly. For instance,
$\varepsilon^{\mathrm{NF}}_{\mu \tau}$ may have some interesting
effects, while $\varepsilon^{\mathrm{NF}}_{\tau \mu}$ is completely
irrelevant, since the beam does not contain any $\nu_\tau$.

The leptonic NSI effects in matter are only sensitive
to the vector component as
\begin{eqnarray}
\varepsilon^{m}_{\beta \alpha} =  (\varepsilon^{L})^{\alpha
e}_{\beta e} +  (\varepsilon^{R})^{\alpha e}_{\beta e} \, ,
\label{equ:matter}
\end{eqnarray}
where two charged leptons are restricted to electrons. From
Eqs.~\eqref{equ:nsi} and \eqref{equ:matter}, we find that
$\varepsilon^m_{\alpha \beta}=(\varepsilon^m_{\beta \alpha})^*$. In
addition, since the beam typically transverses ordinary matter
consisting of electrons, these parameters are (almost) experiment
independent.

Now, if we apply the antisymmetric condition (from gauge invariance
and no charged-lepton-flavor-violation), the matter NSI parameters
$\varepsilon^m_{e\alpha}$ (for $\alpha=e,\mu,\tau$) and the source
NSI parameters $\varepsilon^{\mathrm{NF}}_{e\mu}$ and
$\varepsilon^{\mathrm{NF}}_{\mu e}$ are forbidden. In addition, we
have the following relations~\cite{Gavela:2008ra}
\begin{eqnarray}
\label{equ:rel0} \varepsilon^m_{\mu \mu} & = &  -
\varepsilon^{\mathrm{NF}}_{ee} = - \varepsilon^{\mathrm{NF}}_{\mu
\mu} \, ,  \\ \varepsilon^m_{\mu \tau} & = &  - (
\varepsilon^{\mathrm{NF}}_{\mu \tau})^* \, , \label{equ:rel}
\end{eqnarray}
with both $\varepsilon^m_{\tau \tau}$ and
$\varepsilon^{\mathrm{NF}}_{e \tau}$ being uncorrelated. Of course,
if charged lepton flavor violation is only suppressed, the relations
in \equ{rel} only hold to some degree. However, we assume that the
underlying theory does not produce charged-lepton-flavor-violation,
such as the mentioned singly charged scalar.

\subsection{Dimension-six operators mediated by fermions}

In general, gauge invariant theories extended the SM with the
tree-level exchange of heavy neutral fermions result in a
dimension-six operator in the form of \equ{LNU}
\cite{Abada:2007ux,Antusch:2006vwa}
\begin{eqnarray}
{\cal O^F} = c_{\alpha\beta} \left(\overline{L}_\alpha
\tilde\phi\right) {\rm i} \slc\partial \left(\tilde\phi^\dagger
L_\beta\right) \, , \label{eq:LNU}
\end{eqnarray}
with $c_{\alpha\beta}$ being the model dependent coefficients and
$\tilde\phi$ being related to the Higgs doublet by $\tilde\phi =
{\rm i} \tau_2 \phi^*$.\footnote{Apart from fermion singlets, this
operator could also be realized in other frameworks once
cancellations or loop-induced effects are taken into account.} After
spontaneous breaking of the SM gauge symmetry, the operator defined
in Eq.~\eqref{eq:LNU} leads to a correction of the neutrino kinetic
energy term, and hence, the leptonic mixing matrix deviates from
unitarity. Note that the NU effects only make sense by means of
effective theories, while unitarity will be restored once the
``full" theory is taken into account.

In the case of a non-unitary leptonic mixing matrix, the mass
eigenstates (the physical states) of neutrinos are linked to their
flavor eigenstates by means of a non-unitary transformation
\cite{Altarelli:2008yr}
\begin{eqnarray}\label{eq:N}
|\nu_f \rangle = N |\nu_m \rangle  = (1+\eta)U |\nu_m \rangle\, ,
\end{eqnarray}
where $\eta \simeq - c v^2 / 2$ is a Hermitian matrix and $U$ is a
unitary matrix diagonalizing the neutrino mass matrix. Note the
similarity to Eqs.~(\ref{eq:s}) and~(\ref{eq:d}) with respect to the
source and detector effects; however, one can also see the
difference compared to NSIs: for both NSIs and NU the matter effects
are given in terms of $|\nu_m \rangle $, which, however, appear on
the left-hand-side of Eq.~(\ref{eq:N}) and implicitly in
Eq.~(\ref{eq:s}). In the NU case, the flavor basis, through which
the NC and CC interactions are defined, is slightly shifted by
$\eta$. In the NSI case, additional contributions at source and
detector may be present, which do not necessarily affect the
properties of the weak interactions in matter (which are still
defined with respect to the original flavor eigenstates, and the
link between mass and flavor eigenstates remains unitary). The time
evolution of neutrino flavor eigenstates is given by
\begin{eqnarray}
{\rm i} \frac{\rm d}{{\rm d}t} |\nu_f \rangle  = \left(1+\eta\right)
{\cal H} \left(1+\eta\right)^{-1} |\nu_f \rangle \, ,
\end{eqnarray}
where
\begin{eqnarray}
{\cal H} & = & {\cal H}_0 + \left(1+\eta\right)^\dagger {\rm diag}
\left(V_{\rm CC}-\frac{1}{2}V_{\rm NC},-\frac{1}{2}V_{\rm
NC},-\frac{1}{2}V_{\rm NC}\right)
\left(1+\eta\right) \nonumber \\
& = & {\cal H}_0 + {\cal H}_m + {\cal H}_{\rm NC} +
\left\{\left({\cal H}_m+{\cal H}_{\rm NC}\right),\eta\right\} + \eta
({\cal H}_m+{\cal H}_{\rm NC}) \eta \, \nonumber \\
& \simeq & {\cal H}_0 + {\cal H}_m + \left\{\left({\cal H}_m+{\cal
H}_{\rm NC}\right),\eta\right\}  + {\cal O}(\eta^2)\, .
\label{eq:HUV}
\end{eqnarray}
Here ${\cal H}_{\rm NC} = -\frac{1}{2}{\rm diag} \left(V_{\rm
NC},V_{\rm NC},V_{\rm NC}\right)$ and $V_{\rm NC} \simeq \sqrt{2}G_F
N_n $, with $N_n$ being the number density of neutrons in Earth
matter. Since NU effects are sub-leading order effects, one can
safely neglect the terms proportional to $\eta^2$ in
Eq.~\eqref{eq:HUV}. The pure NC contribution [the third term in the
second row of Eq.~\eqref{eq:HUV}] is flavor blind, and hence can
also be dropped. Then, by comparing Eq.~\eqref{eq:HUV} with
Eq.~\eqref{equ:H}, we obtain the relations \cite{Antusch:2008tz}
\begin{eqnarray}
{\cal H}_{\rm NSI} & = & \left\{\left({\cal H}_m+{\cal H}_{\rm
NC}\right),\eta\right\} \, , \\
\varepsilon^m_{\alpha\beta} & = & \eta_{\alpha e} \delta_{\beta e}+
\eta_{e\beta} \delta_{\alpha e} - \frac{V_{\rm NC}}{V_{\rm CC}}
\eta_{\alpha\beta} \, ,
\label{equ:nccc}
\end{eqnarray}
for neutrino propagation in matter, and
\begin{eqnarray}
\label{eq:muv1}
\varepsilon^s_{\alpha\beta} = \varepsilon^d_{\alpha\beta} =
\eta_{\alpha\beta} \, ,
\end{eqnarray}
for the source and detector effects. Note that, for neutrino
propagation in realistic Earth matter, $N_e \simeq N_n$ holds to a
very good precision. Therefore, we have approximately
\begin{eqnarray}
\varepsilon^m_{ee} & \simeq &2 \eta_{ee} \, , \quad
\varepsilon^m_{\mu \mu} \simeq - \eta_{\mu \mu} \, , \quad
\varepsilon^m_{\mu \tau}\simeq - \eta_{\mu \tau} \, , \quad \,
\varepsilon^m_{\tau\tau} \simeq-\eta_{\tau\tau} \, ,
\label{eq:relMUV}
\end{eqnarray}
together with $\varepsilon^m_{e\mu} \simeq \varepsilon^m_{e
\tau}=0$. In practice, these cancellations only hold up to the
percentage level, depending on the composition of the
material~\cite{Antusch:2008tz}. This implies that
$\varepsilon^m_{e\mu}$ and $\varepsilon^m_{e \tau}$ are expected to
be suppressed by two orders of magnitude compared to the other
parameters.

In comparison with the NSI effects in the previous subsection, a
salient feature is that, both the source and detector effects exist,
and they are process and experiment independent. In the mean while,
they always lead to interferences between the non-standard effects
and the standard oscillation effects.

\subsection{Other categories of non-standard effects}

Apart from purely leptonic NSIs, there could be NSI operators
involving quarks, or NSIs from leptonic dimension-six operators,
which do lead to charged-lepton-flavor-violation. Furthermore, NSIs
may be induced by dimension-eight or higher operators. For
dimension-eight operators at tree level, however, ${\cal O}^6$
effects (either NSIs or NU or both) are induced as well, or exotic
fermions appear, which are strongly constrained by electroweak
precision tests~\cite{Antusch:2008tz,Gavela:2008ra}. There exists
the principle possibility that the dimension-six NSI operators
coming from different sources cancel and the leptonic NSI effects
originate exclusively from dimension-eight or higher
operators~\cite{Gavela:2008ra}. In such a case, the NSI operators
might not produce charged lepton flavor violation either. The NSI
operators involving four lepton doublets [second category in
\equ{opnsi}] allow then for a connection between source and matter
effects, which could be different from the one discussed in
\Sec~\ref{sec:scalarvector}. On the other hand, if NSIs come from
operators with two lepton doublets and two singlets [first category
in \equ{opnsi}], only matter NSIs will be
induced~\cite{Gavela:2008ra}.

In summary, there is a third category of non-standard effects, for
which the source, detector, and matter interactions may be
uncorrelated or correlated in a different way than in the previous
subsections. However, these possibilities are either suppressed by
higher orders of the new physics scale (dimension-eight operators),
or face other constraints. For the sake of simplicity, we do not
discuss these categories in detail any further, but we will in some
cases point out when observations correspond to this category
``Other''. One should also keep in mind that data can only be
interpreted in certain ways induced by new physics. We mainly
discuss the interpretation in terms of dimension-six operators.

\subsection{Bounds on the dimension-six operators}
\label{sec:bounds}

The bounds on the dimension-six operators are, in fact, model
dependent. The experimental bounds mainly come from the lepton
flavor violating decays $\ell_\alpha \to \ell_\beta \gamma$, the
universality test of weak interactions and the invisible decay width
of the $Z$-boson and have been summarized in
Ref.~\cite{Antusch:2008tz}. For example, for a scalar mediated
${\cal O^S}$ operator, one has
\begin{eqnarray}
|\varepsilon^m_{\mu\mu}|  =
|\varepsilon^{\mathrm{NF}}_{ee}|=|\varepsilon^{\mathrm{NF}}_{\mu\mu}|
& < & 8.2\times10^{-4}
\, , \\
|\varepsilon^m_{\mu\tau}|  = |\varepsilon^{\mathrm{NF}}_{\mu\tau}| & < &
1.9\times10^{-3} \, ,\\
|\varepsilon^m_{\tau\tau}| & < & 8.4 \times10^{-3} \, , \\
|\varepsilon^{\mathrm{NF}}_{e\tau}| & < & 7.5 \times10^{-2} \, .
\end{eqnarray}
For the non-unitarity effects induced by the ${\cal O^F}$ operator,
if the mediators are heavier than the electroweak scale, one has
upper bounds on the $\eta$ parameters\footnote{Note that, compared
with the bounds on $NN^\dagger$ in Ref.~\cite{Antusch:2008tz}, the
constraints on $\eta$ are strengthened by a factor 2, since
$NN^\dagger \simeq 1+2\eta$ according to Eq.~\eqref{eq:N}.}
\begin{eqnarray}
|\eta| < \left(\begin{matrix} 2.0 \times 10^{-3}  & 6.0 \times
10^{-5}  & 1.6 \times 10^{-3}  \cr \sim & 8.0 \times 10^{-4}  & 1.1
\times 10^{-3} \cr \sim & \sim & 2.7\times 10^{-3}
\end{matrix} \right) \, .
\end{eqnarray}
If the mediators are lighter than the electroweak scale but above a
few GeV, the above bounds still apply except
$|\eta_{e\mu}|<9.0\times 10^{-4}$ has to be employed because of the
restoration of unitarity in the $Z$-decay. In the case that NSIs
come exclusively from $d\geq 8$ NSI operators, the severe
constraints from universality test may not apply coherently, and
hence, the bounds on NSI parameters are rather loose.

The more general, model independent NSI bounds for a neutrino
factory (at 90~\% C.L.) are given by~\cite{Biggio:2009nt}
\begin{eqnarray}
|\varepsilon^{\mathrm{NF}}_{\alpha\beta}| < \left(\begin{matrix}
0.025 & 0.030 & 0.030  \cr 0.025 & 0.030 & 0.030 \cr 0.025 & 0.030 &
0.030
\end{matrix} \right) \, ,
\end{eqnarray}
and
\begin{eqnarray}
|\varepsilon^{mL}_{\alpha\beta}| < \left(\begin{matrix} 0.06 & 0.10
& 0.4  \cr \sim & 0.03 & 0.10 \cr \sim & \sim & 0.16 \end{matrix}
\right) \, , \ \ \ \ \ |\varepsilon^{mR}_{\alpha\beta}| <
\left(\begin{matrix} 0.14 & 0.10 & 0.27  \cr \sim & 0.03 & 0.10 \cr
\sim & \sim & 0.4
\end{matrix} \right) \, ,
\end{eqnarray}
for left- and right-handed NSIs, respectively.

From this comparison, one can already see one dilemma which could be
called the ``NSI hierarchy problem''. While the model-independent
bounds are relatively weak compared to the bounds on the
dimension-six operators, the theory leading to such large
non-standard effects cannot be that straightforward. For instance,
if the NSIs came from dimension-eight operators, they would be
naturally expected to be of the order $v^4/\Lambda_{\mathrm{NP}}^4
\simeq 10^{-4}$ for $\Lambda_{\mathrm{NP}} =
\mathcal{O}(1)~\mathrm{TeV}$ (as expected by the hierarchy problem).
However, in this case, one cannot use particular correlations
between source and matter NSIs to enhance the sensitivity. This
means that the non-standard effects may in either case be on the
edge of the sensitivity of a neutrino factory (see, \eg,
\Ref~\cite{Kopp:2008ds} for matter NSIs).

\section{Testing the origin of non-standard effects}
\label{sec:test}

\begin{table}[t]
\begin{center}
\begin{tabular}{|c|c|c|c|c|c|c|c|c|c|c|c|c|c|c|c|c|c|c|c|}
\hline  & \multicolumn{2}{|c|}{$\nu$-factory} &
\multicolumn{2}{|c|}{SB} && \multicolumn{2}{|c|}{$\nu$-factory} &
\multicolumn{2}{|c|}{SB} && \multicolumn{2}{|c|}{$\nu$-factory} &
\multicolumn{2}{|c|}{SB} \\
\cline{2-5} \cline{7-10} \cline{12-15} & ${\cal O^S}$ & ${\cal O^F}$
& ${\cal O^S}$& ${\cal O^F}$ & & ${\cal O^S}$ & ${\cal O^F}$ &
${\cal O^S}$ & ${\cal O^F}$ & & ${\cal O^S}$ & ${\cal O^F}$ & ${\cal
O^S}$ & ${\cal O^F}$
\\
\hline  $\varepsilon^m_{ee}$ &   & \ding {52} &   & \ding {52} & $\varepsilon^s_{ee}$ & \ding {52} & \ding {52} & n/a & n/a &&&&& \\
$\varepsilon^m_{e\mu}$ &   &  &  & & $\varepsilon^s_{e\mu}$ & & \ding {52} & n/a & n/a &&&&&\\
 $\varepsilon^m_{e\tau}$ &   &  &   & & $\varepsilon^s_{e\tau}$ & \ding {52} & \ding {52} & n/a & n/a & $\varepsilon^d_{\alpha\beta}$ & & \ding {52} & & \ding {52} \\
 $\varepsilon^m_{\mu\mu}$ & \ding {52}  & \ding {52} &  \ding {52}  & \ding {52} & $\varepsilon^s_{\mu e}$ &  & \ding {52} &  &  \ding {52}& & & & & \\
 $\varepsilon^m_{\mu\tau}$ &  \ding {52} & \ding {52} &  \ding {52}  & \ding {52} & $\varepsilon^s_{\mu\mu}$ & \ding {52} & \ding {52} &  & \ding {52} & & & & & \\
$\varepsilon^m_{\tau\tau}$ &  \ding {52} & \ding {52} &  \ding {52}  & \ding {52} & $\varepsilon^s_{\mu\tau}$ & \ding {52} & \ding {52} & & \ding {52} & & & & & \\
\hline
\end{tabular}
\end{center}
\caption{\it Allowed parameters from the discussed dimension-six
effective operator classes in a neutrino factory ($\nu$-factory) and
a superbeam experiment (SB). \label{tab:allowedNSIs}}
\end{table}

In this section, we qualitatively discuss how the origin of the
non-standard effect can be determined. In the rest of this work, we
then focus on one particular example -- $\varepsilon_{\mu \tau}$ at a neutrino factory.

As we have shown, no matter of their possible origins, the
non-standard effects can always be re-parametrized in terms of
$\varepsilon^s$, $\varepsilon^d$, and $\varepsilon^m$. Therefore, we
have treated them as independent parameters in
\Sec~\ref{sec:formulism}, which can be used for any category.
However, for a given experiment (such as a neutrino factory),
$\varepsilon^m$ will be a particular function of $\varepsilon^s$ in
different frameworks, and not all the NSI parameters are allowed
[\cf, discussion around Eqs.~\eqref{equ:rel0}, \eqref{equ:rel}, and
\eqref{eq:relMUV}]. We summarize in Tab.~\ref{tab:allowedNSIs} which
non-standard effects are allowed for a neutrino factory and a
superbeam if the origin are the discussed leptonic dimension-six
operators.

At this point, we want to emphasize again that the relationships
between source and matter effects in Eqs.~\eqref{equ:rel0},
\eqref{equ:rel}, and \eqref{eq:relMUV} are very similar, especially
for the effects easiest to access. Consider, for instance,
\begin{eqnarray}
\varepsilon^m_{\mu \tau}  & = &    - (\varepsilon^{\mathrm{NF}}_{\mu
\tau})^*  \quad \quad\quad  \text{(NSIs)} \, ,
\label{equ:corr1}\\
\varepsilon^m_{\mu \tau} & = &  - \eta_{\mu \tau} =
-\varepsilon^s_{\mu\tau} \quad \text{(NU)} \, . \label{equ:corr2}
\end{eqnarray}
The origin of these relationships is very different. The NSI
relationship relies on the degree that charged lepton flavor
violation is suppressed, whereas the NU relationship relies on the
equality of the CC and NC potentials in \equ{nccc}.

If one wants to distinguish the origin of the non-standard effects,
it is not sufficient to determine the phase of $\varepsilon^s_{\mu
\tau}$, since one can always fit the data with NSIs with one phase
or NU with minus this phase. In principle, one needs an independent
test of $\varepsilon^s$ and $\varepsilon^m$ (including phases).
However, we will demonstrate that the equality between source and
detector effects for NU leads to additional differences between the
effects.

From the previous discussions, we have qualitatively four
possibilities to determine the origin of the non-standard effects
(\cf, Tab.~\ref{tab:allowedNSIs}):
\begin{itemize}
\item
Distinguish by appearance of certain parameters. If, for instance,
$\varepsilon^m_{ee}$ is found, it cannot be interpreted as
$\mathcal{O^S}$, but as $\mathcal{O^F}$ (or the category ``Other'').
If, on the other hand, $\varepsilon^m_{e \tau}$ is found, it has to
come from the category ``Other'', such as a higher-dimensional
operator. See, \eg, \Ref~\cite{Kopp:2008ds}, for the bounds expected
for these parameters at a neutrino factory. While $\varepsilon^m_{e
\tau}$ is one of the most discussed NSI parameters in the literature
and can be relatively well constrained, $\varepsilon^m_{ee}$ adds to
the standard matter effect and the bounds are limited to the
precision by which the matter density profile is
known~\cite{Kopp:2008ds}. Another such parameter is
$\varepsilon^m_{e\mu}$, which can only come from ``Other'', and is
discussed in \Ref~\cite{Kopp:2007mi}. In addition, $\varepsilon^s_{e
\mu}$ or $\varepsilon^s_{\mu e}$ could discriminate between
$\mathcal{O^S}$ and $\mathcal{O^F}$. At a neutrino factory near
detector, this measurement will be limited by charge identification.
\item
Distinguish by bounds (\cf, \Sec~\ref{sec:bounds}). If large enough
effects beyond the dimension-six operator bounds are found, but
below the generic bounds, they have to come from the category
``Other''. If, for instance, $\varepsilon_{e \tau}^{\mathrm{NF}}
\sim 10^{-2}$ is measured, it could come from $\mathcal{O^S}$, but
is excluded for $\mathcal{O^F}$.
\item
Distinguish by experiment class. Since the source NSIs are
production process-dependent parameters and the NU parameters are
fundamental, using a different experiment class can help to
disentangle the effects. For instance, if
$\varepsilon^{\mathrm{NF}}_{\mu \tau}$ is found at the near detector
of a neutrino factory, it may come from $\mathcal{O^S}$ or
$\mathcal{O^F}$. If it is a fundamental parameter from
$\mathcal{O^F}$, it has to appear at a corresponding superbeam near
detector, such as the anticipated Main Injector Non-Standard
Interactions Search (MINSIS) project at Fermilab
(USA)~\cite{MINSIS}. Otherwise, $\mathcal{O^F}$ could be excluded.
Therefore, in order to disentangle process-dependent NSIs from
fundamental NU, neutrino factory and superbeam near detectors are
complementary.
\item
Distinguish by correlations. For example, $\varepsilon^s_{\mu \tau}
= \varepsilon^d_{\mu\tau} = - \varepsilon^m_{\mu\tau}$ holds for
$\mathcal{O^F}$, while $\varepsilon^{\mathrm{NF}}_{\mu \tau} = -
\varepsilon^{m*}_{\mu\tau}$ together with $\varepsilon^d_{\mu\tau} =
0$ holds for $\mathcal{O^S}$. The other correlations in
Eqs.~\eqref{equ:rel0} and \eqref{eq:relMUV} will be less accessible
at a neutrino factory, since either there are no $\nu_\tau$ in the
beam or the measurement  of $\varepsilon^s_{ee}$ and
$\varepsilon^s_{\mu \mu}$ will be intimately connected to the
knowledge of cross sections and fluxes.
\end{itemize}
Since especially the last category requires some more detailed
understanding, we focus on $\varepsilon_{\mu \tau}$ for the rest of
this study. In this case, it has been pointed out in the literature
that the $\nu_\mu \to \nu_\tau$ and $\nu_\mu \to \nu_\mu$ channels
provide us with the best sensitivities (see, \eg,
Refs.~\cite{FernandezMartinez:2007ms,Altarelli:2008yr,Donini:2008wz,Blennow:2008ym,Ota:2002na,Dighe:2007uf,Goswami:2008mi,EstebanPretel:2008qi},
where the phenomenological importance of the $\mu\mu$ and $\mu\tau$
transitions for searching new physics has been stressed. Therefore,
we will mostly concentrate on these two channel in the subsequent
analysis.

\section{Non-standard effects in the $\boldsymbol{\mu}$-$\boldsymbol{\tau}$-sector at a neutrino factory}
\label{sec:last} Here we focus on the measurement of
$\varepsilon_{\mu \tau}$, the differences between the NSI and NU
sensitivities, and the ability to determine the origin of the
non-standard effects at a neutrino factory.

\subsection{Analytical considerations}

In this section, we discuss some of the relevant analytical
properties of the transition probabilities useful to obtain
knowledge on the output of our numerical simulations. We will
concentrate on the appearance probability $P(\nu^s_\mu \rightarrow
\nu^d_\tau)$ as well as the survival probability $P(\nu^s_\mu
\rightarrow \nu^d_\mu)$, keeping all
$\varepsilon_{\alpha\beta}^{s,m,d}$ but
$\varepsilon_{\mu\tau}^{s,m,d}$ vanishing. Note that, since we are
only considering neutrino factory setups, we will drop the upper NF
label on the non-standard parameters. The expressions for these
probabilities can be obtained by applying the general formula given
in Eq.~(\ref{eq:P1}); to further simplify the results, we
consistently expand up to first order in the small quantities
$\theta_{13}$, $\Delta m^2_{21}/\Delta m^2_{31}$, and
$\varepsilon_{\alpha\beta}$. Keeping all the source, matter, and
detector effects, the survival probability reads
\begin{eqnarray}
P_{\mu\mu} &=& 1 - \sin^2 2\theta_{23}\,\sin^2 \left(\frac{\Delta
L}{4 E}\right) - ({\rm Re}\, \varepsilon^d_{\tau\mu}+ {\rm
Re}\,\varepsilon^s_{\mu\tau}) \sin 4\theta_{23}\,\sin^2
\left(\frac{\Delta L}{4 E}\right)   \nonumber \\
\label{pmumugeneral} &&+\left({\rm Im}\, \varepsilon^d_{\tau\mu}+
{\rm Im}\,\varepsilon^s_{\mu\tau}\right)\,
\sin 2\theta_{23}\,\sin \left(\frac{\Delta L}{2 E}\right)  \nonumber \\
&&-{\rm Re}\,\varepsilon^m_{\mu\tau}\sin 2\theta_{23}\left[\frac{a
L}{2 E}\sin^2 2\theta_{23}\,\sin \left(\frac{\Delta L}{2 E}\right) +
\frac{4 a}{\Delta}\cos^2 2\theta_{23}\sin^2 \left(\frac{\Delta L}{4
E}\right)\right] \, ,
\end{eqnarray}
where we defined $a=2EV_{\rm CC}$ and $\Delta =\Delta m^2_{31}$.
This result agrees with Eq.~(35) in Ref.~\cite{Kopp:2007ne}. Note
that the presence of an explicit  CP-violating term in the
disappearance channel is not a signal of breaking the CPT symmetry,
but the result of considering different initial and final states.
This ``CPT violation" is eliminated once the considered production
and detection processes are the same with the same participating
fermion, for which the relation ${\rm Im} \varepsilon_{\tau\mu}^d =
- {\rm Im} \varepsilon_{\mu\tau}^s$ holds. It is now straightforward
to obtain the $\nu_\mu \to \nu_\mu$ transitions in the case of
scalar and fermion mediated dimension-six operators. In the first
case, we need to drop the dependence on $\varepsilon^d_{\mu\tau}$ in
Eq.~(\ref{pmumugeneral}) and use the mapping in Eq.~(\ref{equ:rel});
choosing the source parameter $\varepsilon^s_{\mu\tau}$ as the
relevant parameter, we obtain
\begin{eqnarray}\label{pmumuscalar}
P_{\mu\mu}^{\cal S} &=& 1 - \sin^2 2\theta_{23}\,\sin^2
\left(\frac{\Delta L}{4 E}\right) \nonumber \\
& & -  {\rm Re}\,
\varepsilon^s_{\mu\tau} \sin 4\theta_{23}\,\sin^2 \left(\frac{\Delta
L}{4 E}\right) + {\rm Im}\, \varepsilon^s_{\mu\tau}\,
\sin 2\theta_{23}\,\sin \left(\frac{\Delta L}{2 E}\right) \nonumber  \\
&& +{\rm Re}\,\varepsilon^s_{\mu\tau}\sin 2\theta_{23}\left[\frac{a
L}{2 E}\sin^2 2\theta_{23}\,\sin \left(\frac{\Delta L}{2 E}\right) +
\frac{4 a}{\Delta}\cos^2 2\theta_{23}\sin^2 \left(\frac{\Delta L}{4
E}\right)\right]\,.
\end{eqnarray}
In the case of fermion mediated operators ${\cal O^F}$, we can use
Eqs.~\eqref{eq:muv1} and \eqref{eq:relMUV} to obtain
\begin{eqnarray}\label{pmumufer}
P_{\mu\mu}^{\cal F} &=& 1 - \sin^2 2\theta_{23}\,\sin^2
\left(\frac{\Delta L}{4 E}\right) \nonumber \\
& & - 2\, {\rm Re}\,
\varepsilon^s_{\mu\tau} \sin 4\theta_{23}\,\sin^2 \left(\frac{\Delta
L}{4 E}\right)  \nonumber  \\   &&+ {\rm
Re}\,\varepsilon^s_{\mu\tau}\sin 2\theta_{23}\left[\frac{a L}{2
E}\sin^2 2\theta_{23}\,\sin \left(\frac{\Delta L}{2 E}\right) +
\frac{4 a}{\Delta}\cos^2 2\theta_{23}\sin^2 \left(\frac{\Delta L}{4
E}\right)\right]\, .
\end{eqnarray}
Equations.~(\ref{pmumuscalar}) and~(\ref{pmumufer}) have exactly the
same form, except for the second lines, which are different from
each other. In Eq.~(\ref{pmumufer}), compared to
Eq.~(\ref{pmumugeneral}), it is clear that a cancellation between
the imaginary parts of the source and detector parameters as well as
a sum of their real parts is at work. This is not an accidental
fact, but can be proved to happen for any disappearance probability;
in fact, from the general expression in Eq.~(\ref{eq:P1}), the true
CP violating terms appearing in the last line of Eq.~(\ref{eq:P1})
disappear once we use the definition for ${\cal J}^i_{\alpha\beta}$
in Eq.~(\ref{eq:J}) computed for the same flavor $\alpha = \beta$
and apply the mapping relations given in Eqs.~(\ref{eq:muv1}) and
(\ref{eq:relMUV}).

For the numerical simulations to follow, it is useful to adopt a
{\it short baseline} expansion for the $\nu_\mu \to \nu_\tau$
transition, namely up to second order in small $\varepsilon$'s and
first order in the quantity $\Delta L/ 4 E$. For the general NSI
effects, the transition probability is given by:
\begin{eqnarray}
P_{\mu\tau} &=& \sin^2 2\theta_{23}\,\left(\frac{\Delta L}{4
E}\right)^2 + |\varepsilon^d_{\mu\tau}|^2 + |\varepsilon^s_{\mu\tau}|^2
-2\,\left({\rm Im}\, \varepsilon^d_{\mu\tau}+
{\rm Im}\,\varepsilon^s_{\mu\tau}\right)\,
\sin 2\theta_{23}\,\left(\frac{\Delta L}{2 E}\right) \nonumber \\ &&+ 2\, {\rm Re} \,
(\varepsilon^d_{\mu\tau} \varepsilon^{s*}_{\mu\tau}) \label{pmutaugeneral}
\end{eqnarray}
and it reduces to the following expressions for the scalar and
fermion mediated operators, respectively:
\begin{eqnarray}
\label{pmutauscalar} P_{\mu\tau}^{\cal S} &=&\sin^2
2\theta_{23}\,\left(\frac{\Delta L}{4 E}\right)^2
+|\varepsilon^s_{\mu\tau}|^2- 2\,{\rm
Im}\,\varepsilon^s_{\mu\tau}\,\sin 2\theta_{23}\, \left(\frac{\Delta
L}{2 E}\right) \, , \\ \label{pmutaufer} P_{\mu\tau}^{\cal F}
&=&\sin^2 2\theta_{23}\,\left(\frac{\Delta L}{4 E}\right)^2 +
4\,|\varepsilon^s_{\mu\tau}|^2- 4\,{\rm
Im}\,\varepsilon^s_{\mu\tau}\,\sin 2\theta_{23}\, \left(\frac{\Delta
L}{2 E}\right)\,.
\end{eqnarray}
It is noteworthy that, in these two equations, the third term
dominates over the second if $\varepsilon^s_{\mu\tau} \lesssim
\Delta L/(2 E) \simeq 10^{-3}$ at the energy threshold (about 1~GeV)
of a neutrino factory for $L \simeq 1 \, \mathrm{km}$, whereas the
first term is suppressed by $L^2$. This means that the non-standard
CP violation from ${\rm Im}\,\varepsilon^s_{\mu\tau}$ is, in
principle, measurable in a near detector for large enough
statistics. The typical (relative) statistical error for the near
detectors considered is about $10^{-5}$ to $10^{-6}$, which is to be
compared with $(\varepsilon^s_{\mu\tau})^2$.  In conclusion, the
currently considered detectors are on the edge of measuring that
effect.

\subsection{Simulation techniques}
\label{sec:simu}

We continue to the numerical simulations of searching for different
NSI effects at a  neutrino factory. In our analysis, we mostly
follow the International Design Study (IDS-NF)~\cite{ids} baseline
setup, which consists of $2.5 \times10^{20}$ useful muon decays per
polarity and year with the parent muon energy $E_\mu =25~{\rm GeV}$.
The total running time is assumed to be ten years. Two magnetized
iron calorimeters are assumed at $L=4000~{\rm km}$ (fiducial mass
100 kton) and $L=7500~{\rm km}$ (fiducial mass 50 kton),
respectively. The description of the neutrino factory is based on
\Refs~\cite{ids,Huber:2002mx,Huber:2006wb,Ables:1995wq}. In
addition, we consider in some cases OPERA-inspired (magnetized)
Emulsion Cloud Chamber (ECC) near $\nu_\tau$-detectors. The $\nu_{e}
\to \nu_\tau$ channel description is based on
\Refs~\cite{Autiero:2003fu,Huber:2006wb}. The $\nu_{\mu} \to
\nu_\tau$ channel is assumed to have the same characteristics as in
Ref.~\cite{Autiero:2003fu}, governed by
\Refs~\cite{Donini:2008wz,FernandezMartinez:2007ms}. Since we assume
that the hadronic decay channels of the $\tau$ can be used as well,
we assume a factor of five higher signal and background than in
\Ref~\cite{Autiero:2003fu}, \ie, 48~\% detection efficiency. Because
we are mostly interested in the $\nu_{\mu} \rightarrow \nu_\tau$
channel (for which the $\nu_{e} \rightarrow \nu_\tau$ channel is
only a small perturbation in the presence of $\varepsilon_{\mu
\tau}$ only),  we add the $\nu_\tau$ and $\bar \nu_\tau$ events as a
conservative estimate (however, there is little impact of this
assumption for the parameters considered). The near detector
treatment is based on Ref.~\cite{Tang:2009na} with respect to
geometric effects of decay straight and detector geometry. For the
high-energy neutrino factory, we consider the following near
detectors, based on the simulation described above:
\begin{description}
\item[ND-L]
Large (OPERA-like) size, fiducial mass 2~kt, $d=1~{\rm km}$ (distance to
end of decay straight),
\item[ND-M]
Medium size (\eg, SciBar-sized), fiducial mass 25~t, $d=80~{\rm m}$,
\item[ND-S]
Small size (\eg, silicon vertex-sized), fiducial mass 100~kg,
$d=80~{\rm m}$,
\item[OND@130km]
OPERA-like at intermediate baseline $L=130~{\rm km}$, as proposed in
\Refs~\cite{FernandezMartinez:2007ms,Goswami:2008mi}, in order to
improve the sensitivity to the non-standard effects.
\end{description}
Note that is yet unclear if an ECC can be operated as close as 1~km
to a neutrino factory because of the high scanning load. Therefore,
alternative technologies may be preferable, such as a silicon vertex
detector. In this case, other challenges have to be approached, such
as the background from anti-neutrino charm production. These issues
are currently under discussion within the IDS-NF.

In addition, to a high-energy neutrino factory, we consider a
low-energy version of the neutrino factory based upon
Refs.~\cite{Bross:2007ts,Bross:2009gk}. It has $E_\mu=4.5 \,
\mathrm{GeV}$ and a magnetized Totally Active Scintillator Detector
(TASD) with 20~kt fiducial mass at $L=1300 \, \mathrm{km}$. The
total running time is assumed to be ten years with $7 \times
10^{20}$ useful muon decays per polarity and year. This luminosity
is higher by a factor of 2.8 than the one of the standard neutrino
factory, since all muons are put in one storage ring (factor two)
and the frontend is optimized in a different way, which leads to
another 40~\% increase of the luminosity~\cite{Bross:2009gk}. Note
that we include two types of backgrounds for the appearance
channels, one which scales with the disappearance rates (such as
from charge mis-identification), and one which scales with the
un-oscillated spectrum (such as from neutral current events), both
at the level of $10^{-3}$  (which is a factor of two higher than in
\Ref~\cite{Bross:2009gk}). We also include the $\nu_\mu \rightarrow
\nu_e$ channel, based upon \Refs~\cite{Rubbia:2001pk,Huber:2006wb}.
As near detectors, we consider the above mentioned OPERA-like
detector (ND-L), and the small (ND-S) with the same characteristics
at a distance of 20~m from the end of the decay straight. For the
decay straight, a length of $s=200$~m is assumed~\cite{Tracey}
(needed to compute the effective baseline as described in
\Ref~\cite{Tang:2009na}).

For the experiment simulation, we use the GLoBES
software~\cite{Huber:2004ka,Huber:2007ji} with user-defined
systematics. For the oscillation parameters, we use (see, \eg,
\Refs~\cite{GonzalezGarcia:2007ib,Schwetz:2008er}) $\stheta=0$,
$\sin^2 \theta_{12}=0.3$, $\sin^2 \theta_{23} = 0.5$, $\sdm = 8.0
\times 10^{-5} \, \mathrm{eV^2}$, $\ldm = 2.5 \times 10^{-3} \,
\mathrm{eV^2}$, and a normal mass hierarchy, unless specified
otherwise. We impose external errors on $\sdm$ and $\theta_{12}$ of
3~\% each, and we include a 2~\% matter density
uncertainty~\cite{Geller:2001ix,Ohlsson:2003ip}.

\subsection{Numerical results}

\begin{figure}[t]
\begin{center}\vspace{-0.7cm}
\includegraphics[width=8cm,bb=0 0 720 720]{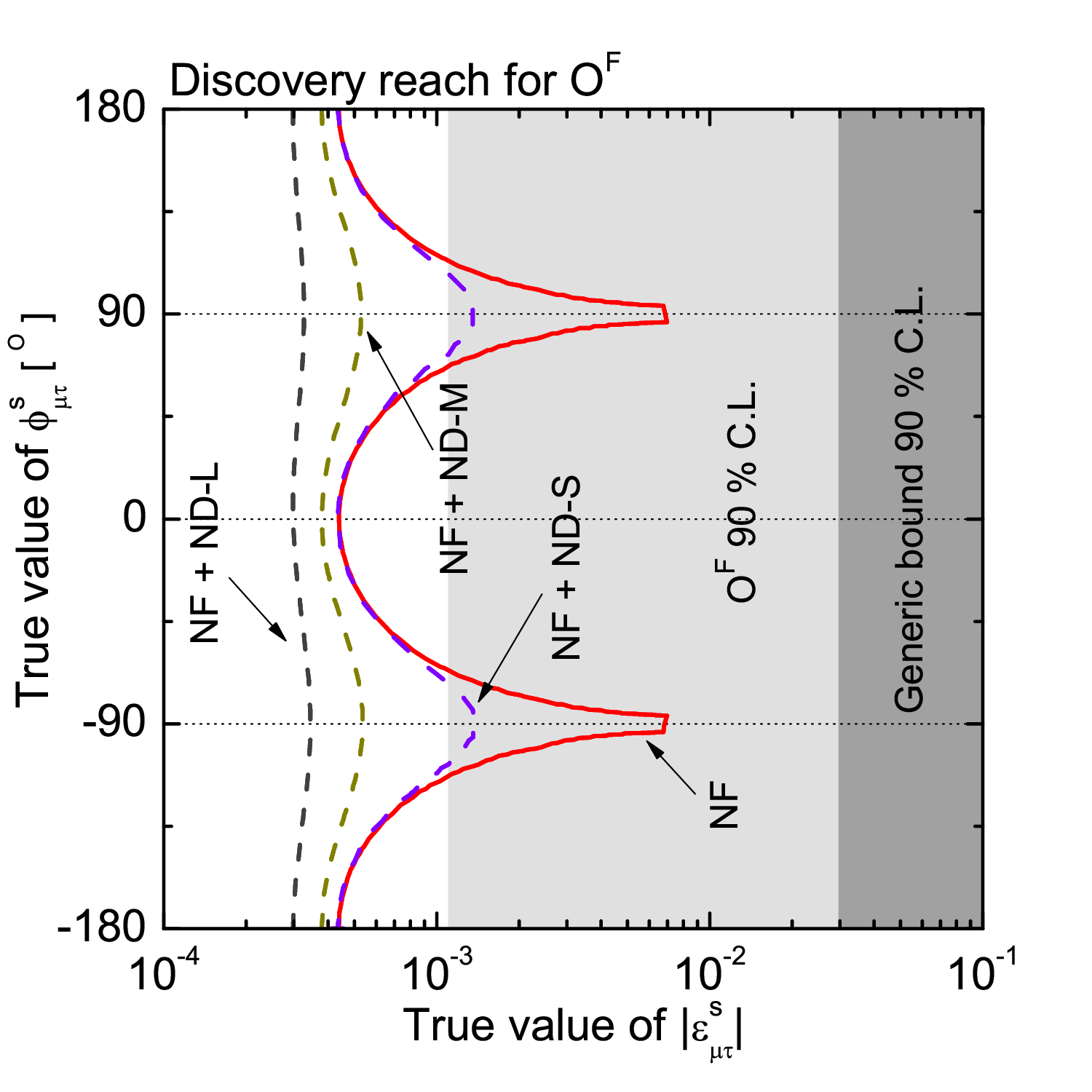}
\includegraphics[width=8cm,bb=0 0 720 720]{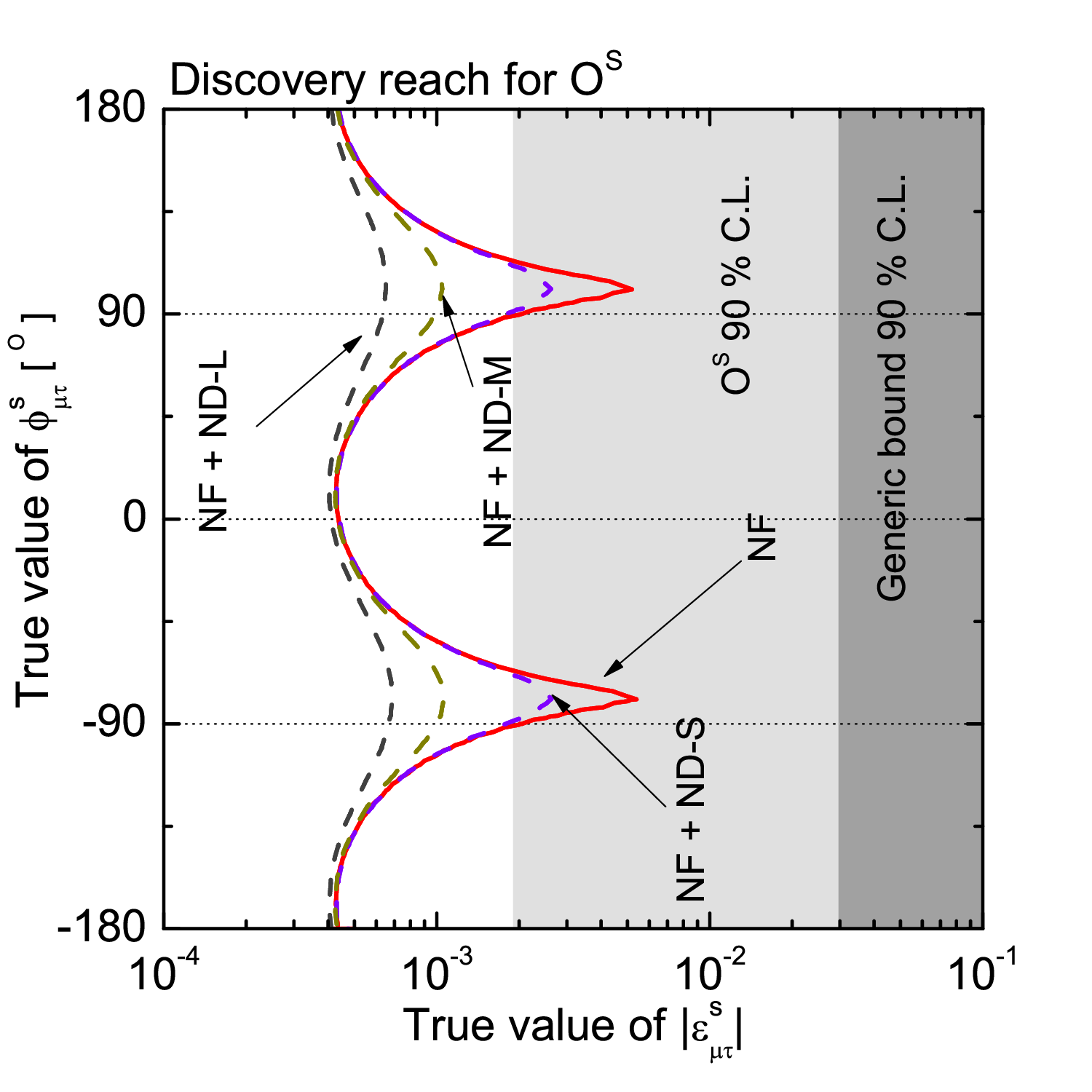}
\vspace{-0.cm}
\caption{\label{fig:fig1}\it  The discovery reach of the IDS-NF
neutrino factory to $\varepsilon^s_{\mu\tau}$ originating from
${\cal O^F}$ (left) or ${\cal O^S}$ (right) operators (on
right-hand-side of curves, 90~\% C.L.). The shadowed areas indicate
the current experimental constraints on the corresponding
dimension-six operators. In both plot the effects of adding
different near detectors are also shown. }
\end{center}
\end{figure}

In Fig.~\ref{fig:fig1}, we illustrate the {\it discovery reach} of
the IDS-NF neutrino factory to the parameter
$\varepsilon^s_{\mu\tau}$ in the two scenarios of fermion (left
panel) and scalar mediated (right panel) operators. We define the
discovery reach as the values of (true) parameters
$|\varepsilon^s_{\mu\tau}|$ and $\phi^s_{\mu\tau}$ [where
$\varepsilon^s_{\mu\tau}=|\varepsilon^s_{\mu\tau}| \exp({\rm i}
\phi^s_{\mu\tau})$], for which $|\varepsilon^s_{\mu\tau}|=0$  can be
excluded at 90~\% C.L. The standard oscillation parameters are
thereby marginalized over.
As we discussed above, in the case of fermionic operators ${\cal
O^F}$, the cancellation between the imaginary parts in the survival
probability leads to the far detectors being sensitive to the real
part of $\varepsilon^s_{\mu\tau}$ only, as it can be clearly seen
from Eq.~\eqref{pmumuscalar}. Therefore, there is hardly sensitivity to
the corresponding $\varepsilon$ parameters in the case of the
CP-violating phase $\phi^s_{\mu\tau} = \pm \pi/2$, as shown by the
solid (red) curve in the left plot of Fig.~\ref{fig:fig1}. In order
to increase the sensitivity to the CP-violating phase
$\phi^s_{\mu\tau}$, it is useful to include the effects of near
detectors capable for $\tau$ identification. In fact, as we can see
from Eq.~\eqref{pmutaufer}, an explicit dependence on the imaginary
part of $\varepsilon^s_{\mu\tau}$ appears in $P_{\mu\tau}^{\cal F}$
and, depending on the type of the near detector, this term could be
relevant. Note also that the $|\varepsilon_{\mu \tau}^s|^2$ term
helps in increasing the sensitivity to the absolute value of
$\varepsilon^s_{\mu\tau}$. In this plot, we also compare three
different scenarios, in which we combine the IDS-NF neutrino factory
setup with the near detectors ND-S, ND-M, and ND-L. The best
combination is using the ND-L detector, mainly due to the larger
mass. As a rough estimate, a sensitivity $3\times10^{-4}$ to
$|\varepsilon^s_{\mu\tau}|$  can be expected in the presence of a
near $\nu_\tau$-detector.  Note also that a short distance
$\nu_\tau$-detector ($L\simeq 100 ~{\rm km}$) does not help much in
improving the sensitivity, and hence, we did not include the
relative curve in the plot. For a scalar mediated NSI operator, the
situation is a bit different, due to the non-vanishing contribution
from the imaginary part of $\varepsilon^s_{\mu\tau}$, see
Eq.~\eqref{pmutauscalar}. Although this term is dominated by the
matter-induced one, its effect is already visible in the right plot
of Fig.~\ref{fig:fig1}, where the CP violating term is responsible
for the asymmetric behavior of the sensitivity curves with respect
to $\phi^s_{\mu\tau}=0$, and better sensitivity for $\phi^s_{\mu
\tau} \simeq \pm 90^\circ$. Similar to the ${\cal O^F}$ case, a
better sensitivity could be expected once the larger near detector
is taken into account (the scenario label NF + ND-L in the plot).

\begin{figure}[t]
\begin{center}\vspace{-0.7cm}
\includegraphics[width=8cm,bb=0 0 720 720]{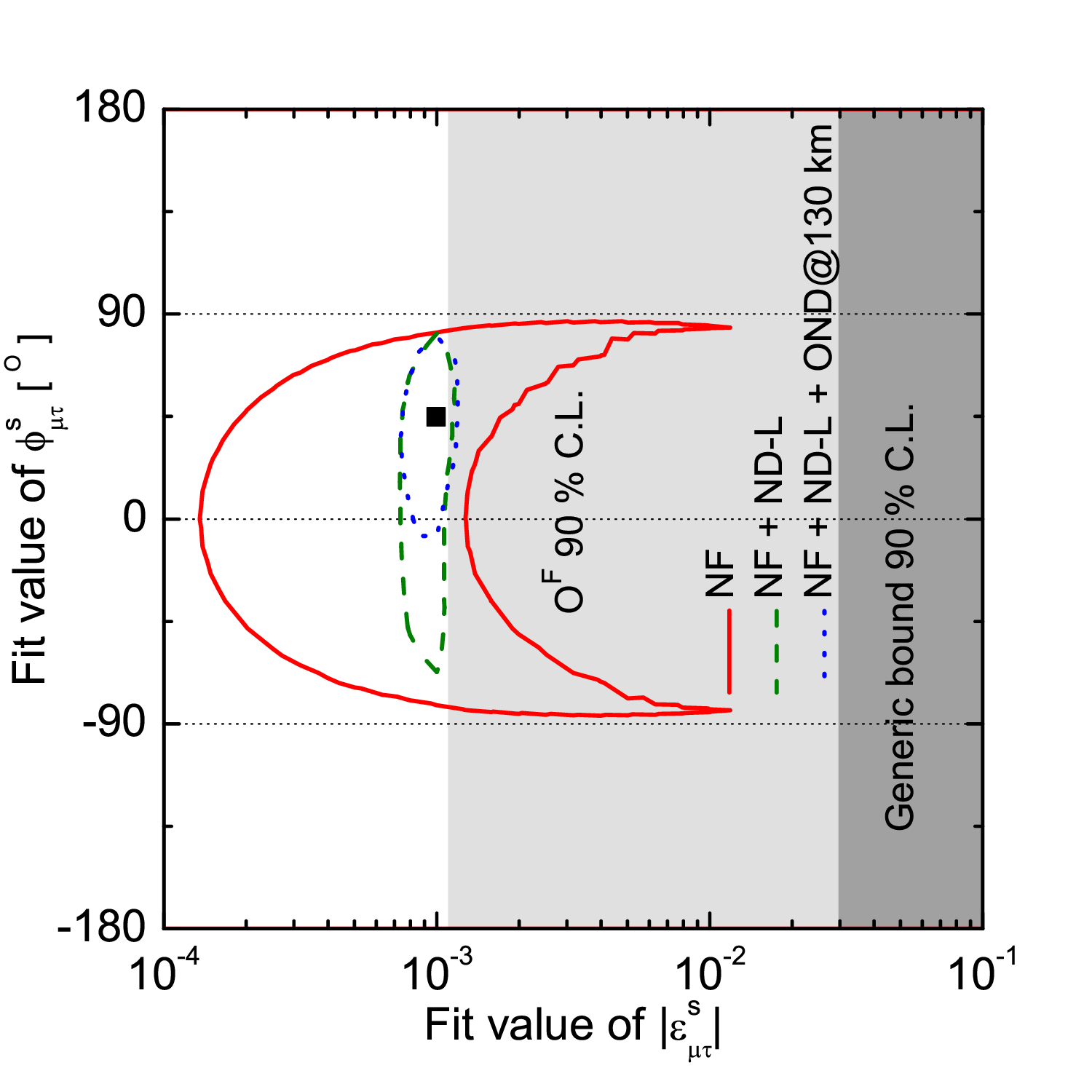}
\includegraphics[width=8cm,bb=0 0 720 720]{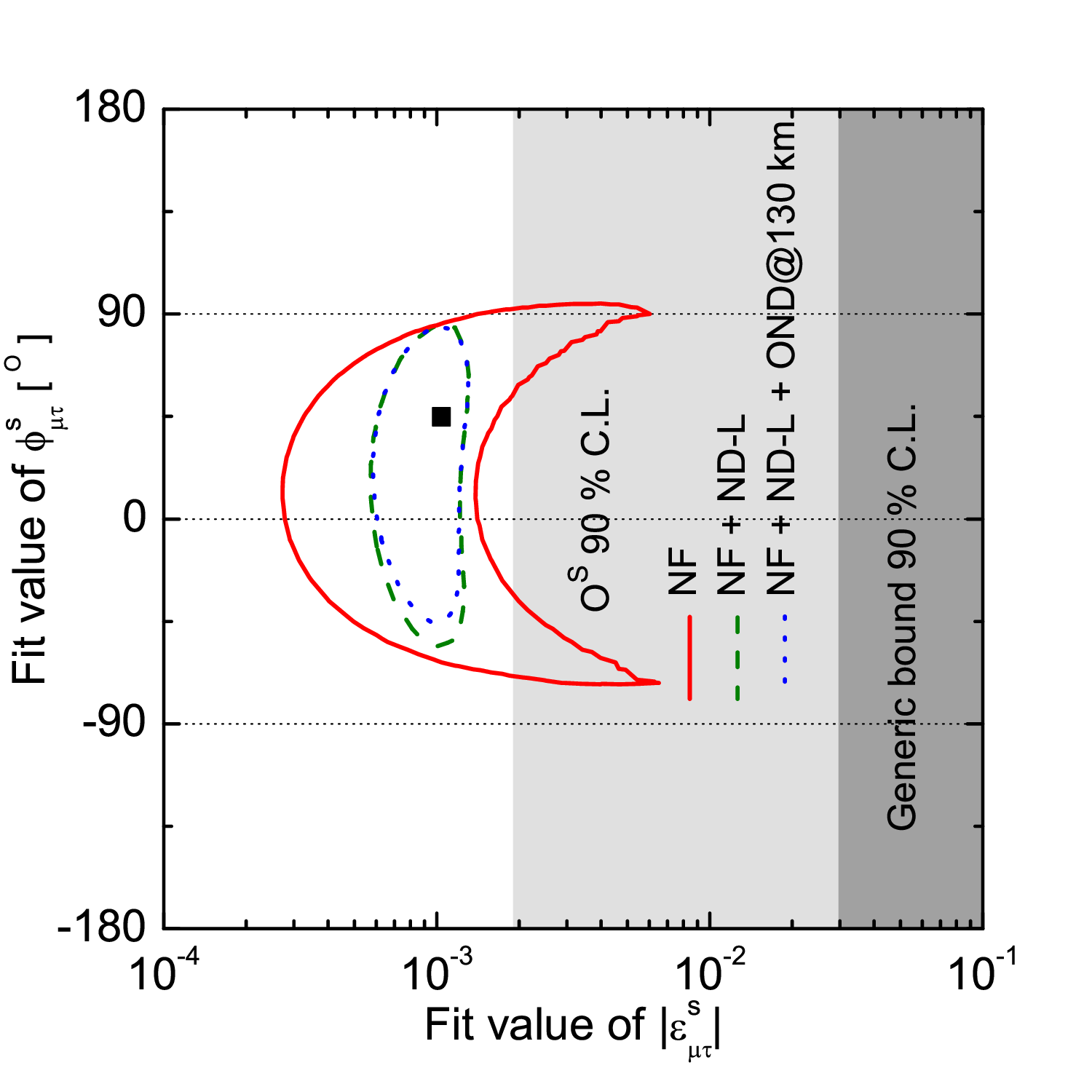}
\vspace{-0.cm}
\caption{\label{fig:fig2}\it The 90~\% C.L. contour plot of the NSI
parameter $\varepsilon^s_{\mu\tau}$ originating from ${\cal O^F}$
(left plot) or ${\cal O^S}$ (right plot), assuming the {\it true}
parameter $\varepsilon^s_{\mu\tau} = 0.001 \exp ({\rm i}\pi/4)$. The
analyzed experimental setups consider the IDS-NF neutrino factory
alone and in combination with an OPERA-like near detector at a
baseline of 1 km and/or 130 km.}
\end{center}
\end{figure}

Compared to standard oscillation physics, where the statistics in
the far detectors limit the performance, the size of the near
detector is very important for non-standard effects. A meaningful
question could be how well the non-standard effects can be measured
if they are not vanishing. An example is given in
Fig.~\ref{fig:fig2}, in which the best-fit contours for the chosen
NSI parameter $\varepsilon^s_{\mu\tau} = 0.001~{\rm exp}({\rm
i}{\pi}/{4})$ are plotted for both type of operators in three
different situations, where the IDS-NF neutrino factory setup  is
alone or accompanied by one or two OPERA-like detectors at different
baselines, namely at $L=1$ km and $L=130$ km. Using the other two
options ND-S and ND-M do not improve the performance.
Due to the fact that the far muon-detector is not sensitive to the
imaginary part of $\varepsilon^s_{\mu\tau}$ (expect from a small
effect for ${\cal O^S}$), we observe in both panels  that a standard
neutrino factory without near detectors has almost no sensitivity to
the CP-violating part $\sin \phi^s_{\mu\tau}$. The situation
improves a lot in the presence of near detectors, especially when a
short-distance detector located at $L\simeq 100~{\rm km}$ is taken
into account, in which the $\sin\Delta$ terms in
Eqs.~\eqref{pmutauscalar} and \eqref{pmutaufer} contribute to the
appearance probability. In particular, the CP-violating phase of
$\varepsilon^s_{\mu\tau}$ can be better reconstructed for the ${\cal
O^F}$ operator (left panel) because of the additional improving
factor of two in the imaginary part term.

\begin{figure}[t]
\begin{center}\vspace{-0.7cm}
\includegraphics[width=8cm,bb=0 0 720 720]{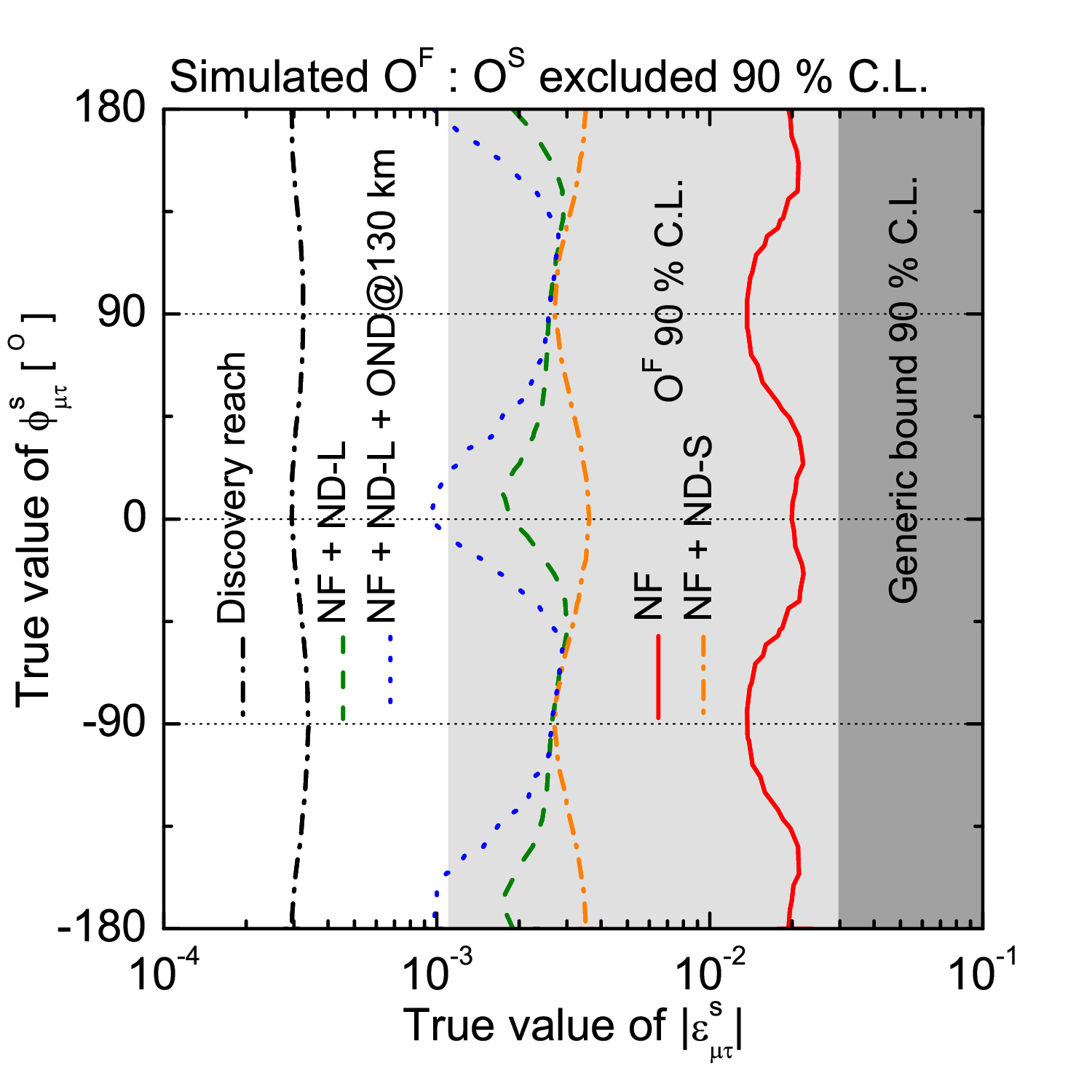}
\includegraphics[width=8cm,bb=0 0 720 720]{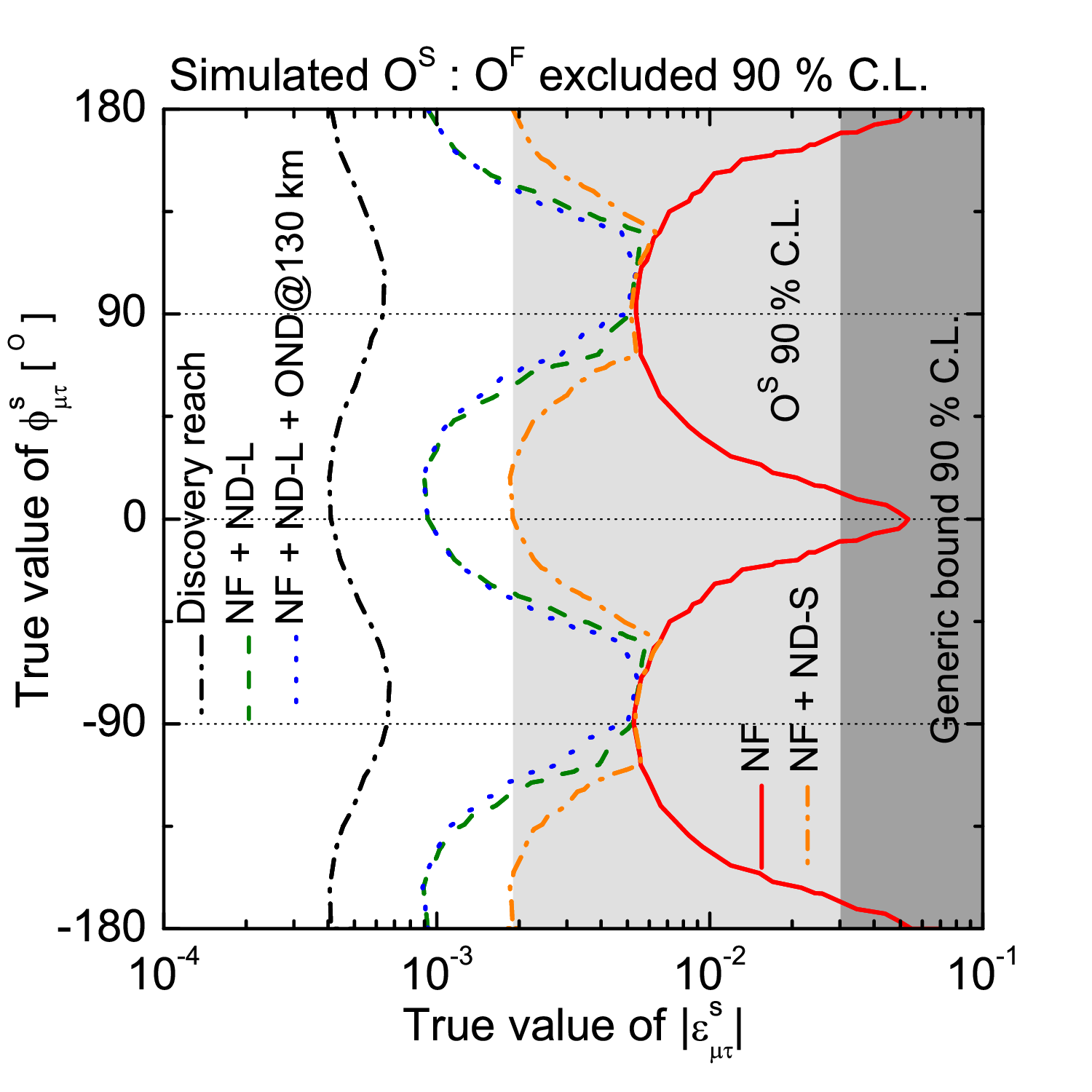}
\vspace{-0.cm}
\caption{\label{fig:fig3} \it Regions in the
($|\varepsilon^s_{\mu\tau}|$-$\phi^s_{\mu\tau}$)-plane where the
simulated $\epsilon^s_{\mu\tau}$ induced by one type of operator can
be uniquely established, \ie, the other type of operator is excluded
at the 90~\% C.L.  (regions on the right-hand side of the curves).
Left panel: the simulated  $\varepsilon^s_{\mu\tau}$ is induced by
${\cal O^F}$ and  fitted with ${\cal O^S}$. Right panel: the
simulated  $\varepsilon^s_{\mu\tau}$ is induced by ${\cal O^S}$ and
fitted with ${\cal O^F}$. In both panels, the discovery reach is
also displayed. The experimental setup is the same as
Fig.~\ref{fig:fig2}.}
\end{center}
\end{figure}

Since the phenomenological signatures  of  ${\cal O^F}$ and ${\cal
O^S}$ in terms of discovery potential and parameter measurements are
quite similar, it is an important question to see if a neutrino
factory-based setup is able to discriminate ${\cal O^F}$ from ${\cal
O^S}$ so as to find hints on the origin of non-standard effects. We
answer this question generating events by using the ``true''
parameters in the case of ${\cal O^F}$ (${\cal O^S}$), and then fit
the data with only ${\cal O^S}$ (${\cal O^F}$). The results of such
an analysis are the exclusion regions (right-hand side of the
curves) shown in Fig.~\ref{fig:fig3}, where the dimension-six
operators can be disentangled.
For the IDS-NF neutrino factory combined with several different near
detectors, the curves in the left panel show that there is just a
very small region beyond the bound at the 90~\% C.L. on the ${\cal
O^F}$ operators, where the data generated with ${\cal O^F}$ can be
distinguished from the ${\cal O^S}$ even if the OPERA-like near
detector at the longer baseline is used. If ${\cal O^S}$ is
simulated, however, it can be distinguished from ${\cal O^F}$ for a
part of the parameter space beyond the current bound with ND-L, as
shown in the right panel of Fig.~\ref{fig:fig3}, especially around
the CP-conserving value $\phi^s_{\mu\tau}=0,\pm \pi$, where the
current experimental constraints on ${\cal O^S}$ are not as
stringent as for ${\cal O^F}$. The combinations with other near
detectors are illustrated by the ND-S curves, where the sensitivity
does not go beyond the current bounds.

From \figu{fig3}, we can read off that there are substantial regions
of the parameter space between the curves, where ${\cal O^S}$ and
${\cal O^F}$ can be distinguished at the neutrino factory itself, and
the discovery reach of the neutrino factory, for which the neutrino
factory will find a non-standard effect, but cannot classify it.
These regions go beyond the current bounds. As mentioned before, an
interesting discriminator might in this case be a superbeam
$\nu_\tau$-detector, such as the MINSIS project. Such a detector
could tell the fundamental effect ${\cal O^F}$ from the
process-dependent effect ${\cal O^S}$ if the sensitivity is
comparable to that of the neutrino factory, \ie, $| \epsilon_{\mu
\tau}^s| \ll 10^{-3}$. For instance, if no effect is seen at MINSIS,
but some effect is detected at the neutrino factory, it could come
from ${\cal O^S}$, but not from ${\cal O^F}$.

Recently, a low-energy neutrino factory  ($E_\mu = 4.5 ~{\rm GeV}$)
has been attracting some attention, for which the same kind of
analysis in Figs.~\ref{fig:fig1} and \ref{fig:fig3} can be repeated.
\begin{figure}[t]
\begin{center}\vspace{-0.7cm}
\includegraphics[width=8cm,bb=0 0 720 720]{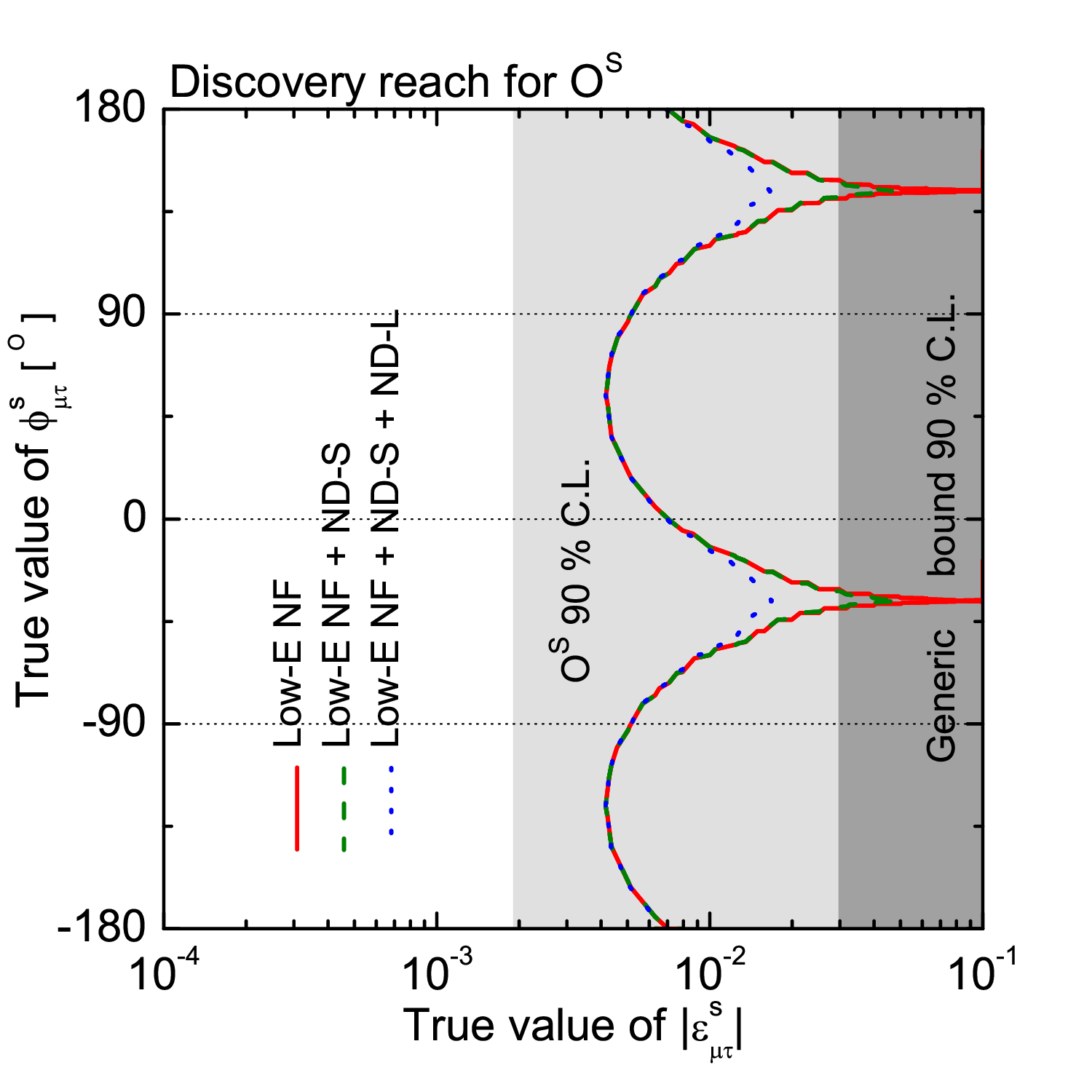}
\includegraphics[width=8cm,bb=0 0 720 720]{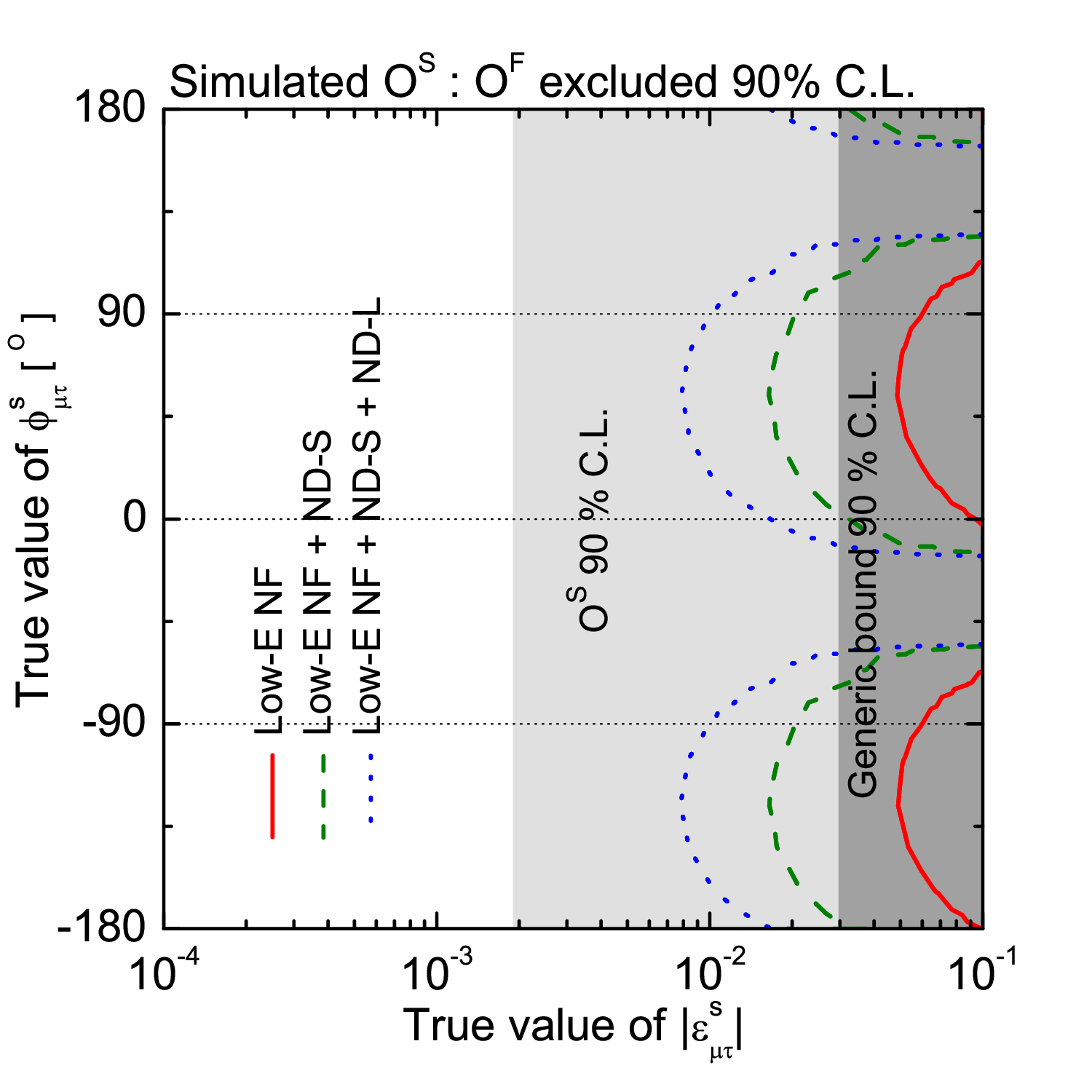}
\vspace{-0.cm}
\caption{\label{fig:fig4} \it Same plots as Fig.~\ref{fig:fig1}
(left) and \ref{fig:fig3} (right), but for a low-energy neutrino
factory alone and in combination with the ND-S and ND-S+ND-L.}
\end{center}
\end{figure}
In the left panel of Fig.~\ref{fig:fig4}, we show the discovery
reach for the ${\cal O^S}$ induced NSIs using the low-energy
neutrino factory alone and in combination with different
combinations of near detectors. We observe that, because of the
$\tau$ production threshold, there is no hope to search for NSIs
originated from  ${\cal O^S}$ operators below the current
experimental limits. Also, in the right plot of Fig.~\ref{fig:fig4}
the same combination of experimental facilities is not able to
exclude the ${\cal O^F}$ operators at 90~\% C.L., since the
exclusion regions are excluded by current limits already. Finally,
in Fig.~\ref{fig:fig5},  we plot the CP discovery potential for both
${\cal O^F}$ (left panel) and ${\cal O^S}$ (right panel) induced CP
violations. This is defined as the ensemble of true values of
$\phi^s_{\mu\tau}$, which cannot be fitted with the CP-conserving
values $\phi^s_{\mu\tau}=0,\pm \pi$ at 90~\% C.L. The combination of
the standard IDS-NF neutrino factory with different large enough
near detectors may discover CP violation, somewhat beyond the
current bounds, especially for $\phi^s_{\mu\tau}\sim \pm \pi/2$.
There are no qualitative differences between ${\cal O^F}$ and ${\cal
O^S}$.

\begin{figure}[t]
\begin{center}\vspace{-0.7cm}
\includegraphics[width=8cm,bb=0 0 720 720]{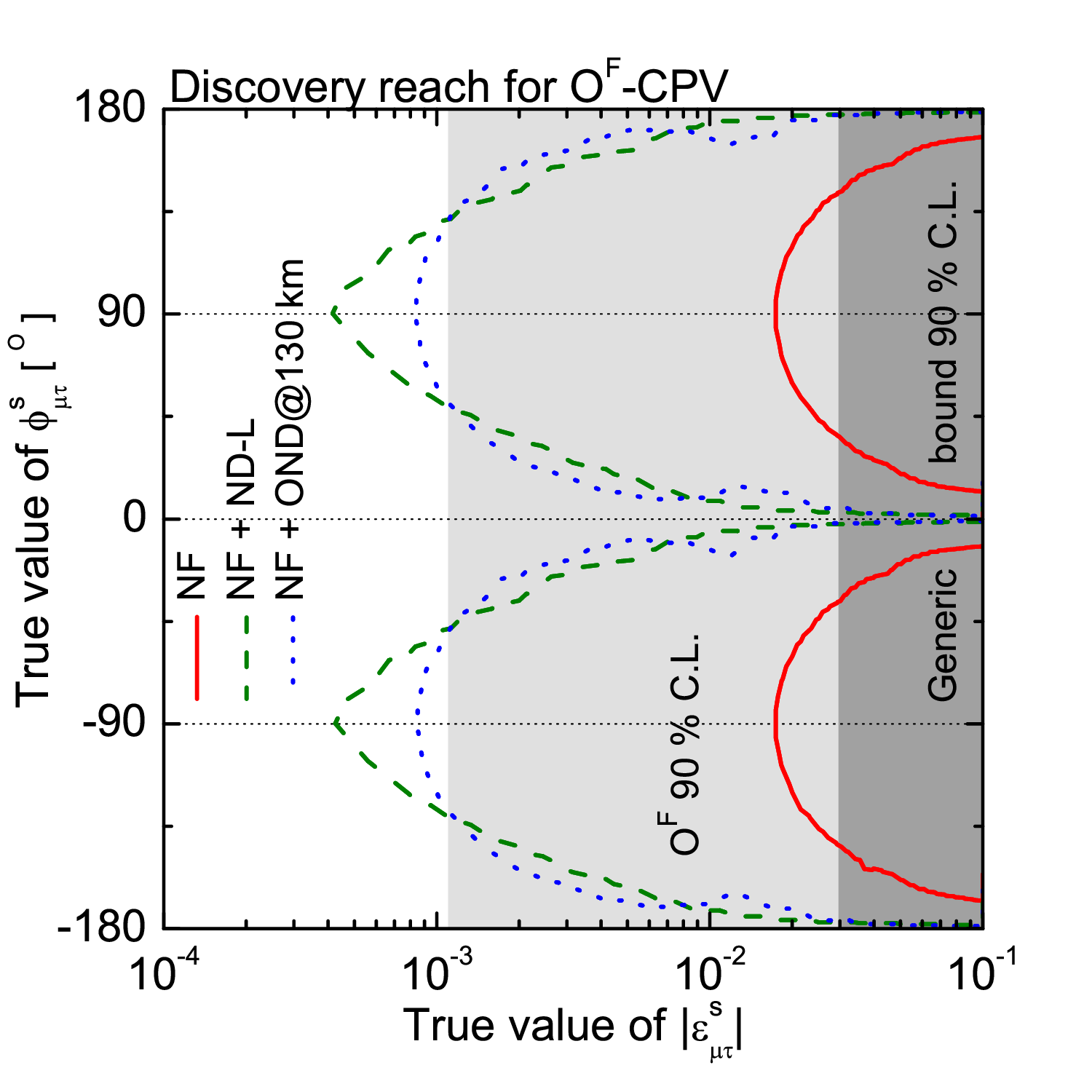}
\includegraphics[width=8cm,bb=0 0 720 720]{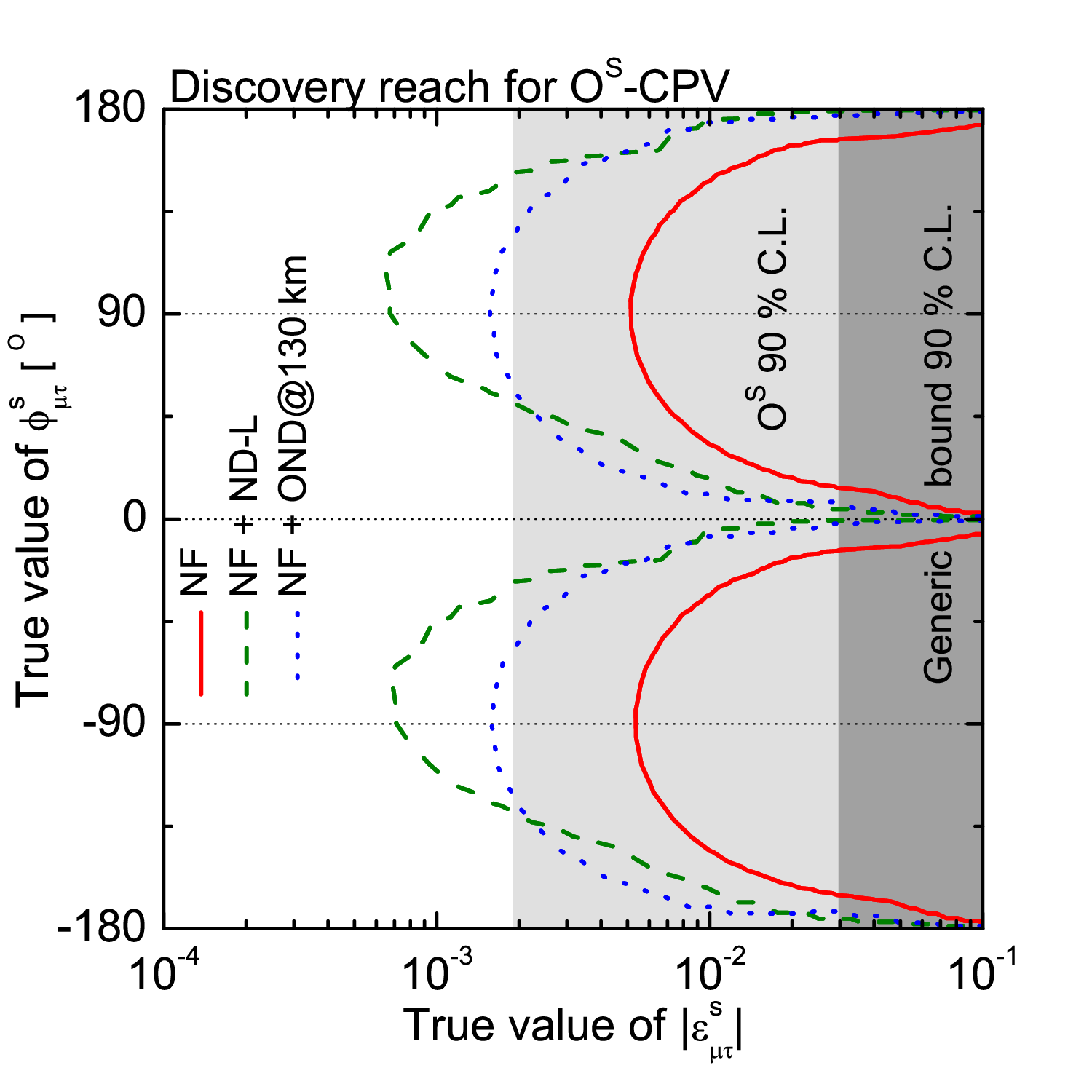}
\vspace{-0.cm}
\caption{\label{fig:fig5}\it The 90~\% C.L. of CP discovery
potentials in the NU framework (left) and NSI framework (right). The
facility combination is the same as Fig.~\ref{fig:fig3}.}
\end{center}
\end{figure}

\section{summary}
\label{sec:summary}

In this work, we have clarified the relationship between NSI and NU
effects, where we have focused on non-standard effects coming from
dimension-six effective operators when the heavy fields are
integrated out. At tree level and without cancellations, these
operators are mediated by scalars (NSIs) or fermions (NU), which
means that the discrimination between NSIs and NU is interesting
from a theoretical point of view, since it may reveal the nature of
the heavy mediator.

From the phenomenological point of view, the assumption of NSIs or
NU, together with the assumptions of gauge invariance and vanishing
charged lepton flavor violation, has lead to particular correlations
between source, propagation, and detector non-standard effects.
These correlations can, in some cases, be used to disentangle NSIs
from NU, relying on the measurement of individual parameters (such
as for $\varepsilon^m_{ee}$). However,  for a neutrino factory, NSIs
and NU look very similar for some parameters -- for entirely
different fundamental reasons.

The most interesting case may be $\varepsilon_{\mu \tau}$, which is,
in principle, easy to find at the near detector of a neutrino
factory (there are no $\nu_\tau$ in the beam). However, the
correlation between source and matter NSIs is basically the same for
NSIs and NU. There is some discrimination potential coming from the
fact that NU has modified source and detector effects, which,
however, hardly exceeds the current bounds. Thus, the easiest way to
discriminate NSIs from NU is the comparison with an experiment using
a different neutrino production mechanism, such as a superbeam. A
possible project in that direction is the MINSIS project. In order
to provide complementary information, a similar sensitivity is
required, which should be significantly below $10^{-3}$ for
$|\varepsilon^s_{\mu \tau}|$.

We conclude that differentiating between NSIs and NU should be one
of the key priorities of searches for new physics effects, since the
nature of the non-standard effect points towards the nature of the
heavy mediator. The components necessary for this search are
$\nu_\tau$ detection at least in near detectors, both at
high-intensity superbeams and a neutrino factory. For the neutrino
factory, a high enough muon energy is mandatory for the discussed
non-standard effects searches, which means that the high-energy
neutrino factory should at least be an upgrade option even for large
$\theta_{13}$. In addition, for non-standard effect searches, the
size of the near detector is very important, which means that for
all applications, large enough detectors are needed.

\begin{acknowledgments}

\vspace{-2.5mm} We wish to thank Toshihiko Ota, Paul Soler, and
Enrique Fern{\'a}ndez-Mart{\'i}nez for useful discussions and
helpful comments. We acknowledge the hospitality and support from
the NORDITA scientific program ``Astroparticle Physics
--- A Pathfinder to New Physics'', March 30 -- April 30, 2009 during
which parts of this study was performed. WW would like to
acknowledge support from the  G{\"o}ran Gustafsson Foundation and
hospitality during his visit in September 2009. This work was
supported by the Royal Swedish Academy of Sciences (KVA) [T.O.], the
G{\"o}ran Gustafsson Foundation [H.Z.], and the Swedish Research
Council (Vetenskapsr{\aa}det), contract no.~621-2008-4210 [T.O.],
and the Emmy Noether program of Deutsche Forschungsgemeinschaft,
contract no. WI 2639/2-1 [D.M. and W.W.]. Furthermore, this work was
supported by the European Union under the European Commission
Framework Programme~07 Design Study EUROnu, Project 212372.

\end{acknowledgments}

\end{document}